\documentclass[12pt, twocolumn]{aastex631}

\usepackage{marginnote}
\received{October 04, 2022}
\revised{December 29, 2022}
\accepted{January 05, 2023}

\submitjournal{ApJ}

\usepackage{tabularx}
\usepackage{amsmath}

\usepackage{color,soul}

\shorttitle{AGN in Post-Merger Remnants}
\shortauthors{Li et al.}

\begin{document}

\title{A Multiwavelength Study of Active Galactic Nuclei in Post-Merger Remnants}

\correspondingauthor{Wenhao Li}
\email{wli58@crimson.ua.edu}

\author{Wenhao Li}
\affiliation{Dept. of Physics and Astronomy, The University of Alabama, Tuscaloosa, AL 35487, USA}

\author{Preethi Nair}
\affiliation{Dept. of Physics and Astronomy, The University of Alabama, Tuscaloosa, AL 35487, USA}

\author{Jimmy Irwin}
\affiliation{Dept. of Physics and Astronomy, The University of Alabama, Tuscaloosa, AL 35487, USA}

\author{Sara Ellison}
\affiliation{Dept. of Physics and Astronomy, The University of Victoria, Victoria, BC V8P 5C2, Canada}

\author{Shobita Satyapal}
\affiliation{Dept of Physics and Astronomy, George Mason University, Fairfax, VA 22030, USA}

\author{Niv Drory}
\affiliation{Dept of Astronomy, The University of Texas at Austin, Austin, TX 78712, USA}

\author{Amy Jones}
\affiliation{Space Telescope Science Institute, Baltimore, MD 21218, USA}

\author{William Keel}
\affiliation{Dept. of Physics and Astronomy, The University of Alabama, Tuscaloosa, AL 35487, USA}

\author{Karen Masters}
\affiliation{Department of Physics and Astronomy, Haverford College, Haverford, PA 19041, USA}

\author{David Stark}
\affiliation{Space Telescope Science Institute, Baltimore, MD 21218, USA}

\author{Russell Ryan}
\affiliation{Space Telescope Science Institute, Baltimore, MD 21218, USA}

\author{Kavya Mukundan}
\affiliation{Dept. of Physics and Astronomy, The University of Alabama, Tuscaloosa, AL 35487, USA}

\begin{abstract}
We investigate the role of galaxy mergers in triggering active galactic nuclei (AGN) in the nearby universe. Our analysis is based on a sample of 79 post-merger remnant galaxies with deep X-ray observations from Chandra/XMM-Newton capable of detecting a low-luminosity AGN of $\ge 10^{40.5}$ erg s$^{-1}$. This sample is derived from a visually classified, volume-limited sample of 807 post-mergers identified in the Sloan Digital Sky Survey Data Release 14 with log M$_*$/M$_{\odot} \ge 10.5$ and $0.02 \le $z$ \le 0.06$. We find that the X-ray AGN fraction in this sample is $55.7\% \pm 5.6\%$ compared to $23.6\% \pm 2.8\%$ for a mass- and redshift-matched noninteracting control sample. The multiwavelength AGN fraction (identified as an AGN in one of X-ray, IR, radio or optical diagnostics) for post-mergers is $76.6\% \pm 4.8\%$ compared to $39.1\% \pm 3.2\%$ for controls. Thus post-mergers exhibit a high overall AGN fraction with an excess between 2 - 4 depending on the AGN diagnostics used. In addition, we find most optical, IR, and radio AGN are also identified as X-ray AGN while a large fraction of X-ray AGN are not identified in any other diagnostic. This highlights the importance of deep X-ray imaging to identify AGN. We find that the X-ray AGN fraction of post-mergers is independent of the stellar mass above log M$_*$/M$_{\odot} \ge 10.5$ unlike the trend seen in control galaxies. 
Overall, our results show that post-merger galaxies are a good tracer of the merger–AGN connection and strongly support the theoretical expectations that mergers trigger AGN.
\end{abstract}
\keywords{Galaxy evolution --- Galaxy interactions --- Galaxy mergers --- Active galactic nuclei}

\section{Introduction}
It is well known that most massive galaxies host a supermassive black hole (SMBH) in their centers \citep{2000ApJ...539L..13G, 2009ApJ...696..891H, 2012ApJ...756..179M, 2013ARA&A..51..511K}. When material is accreted onto a SMBH, it triggers an active galactic nucleus (AGN), which can ionize the gas and/or push gas out of the galaxy (i.e., AGN ``feedback") stopping star-formation. AGN are expected to play a role in regulating the star formation and mass growth of galaxies \citep{2006MNRAS.370..645B, 2006MNRAS.365...11C, 2008ApJS..175..356H, 2015MNRAS.446..521S, 2019MNRAS.490.3196P}. 

Theoretically, mergers of galaxies are expected to trigger AGN \citep{2005ApJ...630..705H,2005ApJ...620L..79S,2008MNRAS.391..481S,2012ApJ...754..125C, 2018MNRAS.481..341S}. The current simplest picture of how bulge-dominated/elliptical galaxies form is: (1) Gas rich, star-forming galaxies merge triggering gas inflows and enhanced star formation in the center. These systems may be observed as (ultra)luminous infrared galaxies (LIRGS). (2) Optical/IR quasars with high Eddington ratios are triggered when gas is accreted onto the SMBHs. (3) The AGN or stellar-driven winds quench star formation by heating and expelling gas, creating post-merger remnants (or bulge-dominated galaxies) with colors that span a range from the blue cloud to the red sequence on the color magnitude diagram. AGN activity declines in this phase until finally (4) a red and dead spheroid is formed, which may exhibit intermittent radio AGN activity \citep{2009ApJ...696..891H}.

Unfortunately, observational support for this scenario is in dispute with some studies finding strong support for mergers triggering AGN while others find no evidence. 
Numerous studies have found a higher optical/mid-IR AGN fraction in galaxy mergers compared to noninteracting galaxies in the local universe. For example, \cite{2011MNRAS.418.2043E} found an optical AGN enhancement of $\sim$2.5 times in Sloan Digital Sky Survey (SDSS) very close pairs ($<$10 kpc) at $0.01 < z < 0.2$ compared to isolated control galaxies. They also found an increasing optical AGN fraction with decreasing projected separation.
Many other low-redshift ($z<0.2$) studies with similar approaches have also found that AGN are more likely to appear in galaxy mergers than isolated galaxies both in the optical \citep{1985AJ.....90..708K, 2007MNRAS.375.1017A, 2007AJ....134..527W, 2013MNRAS.435.3627E} and in the mid-IR \citep{2014MNRAS.441.1297S, 2017MNRAS.464.3882W, 2018PASJ...70S..37G, 2020A&A...637A..94G}. This suggests a connection between AGN and mergers.

Comparatively few studies have probed the X-ray AGN fraction in mergers relative to isolated galaxies in the nearby universe. The recent study presented by \cite{2020MNRAS.499.2380S} found no statistically significant enhancement of X-ray-selected AGN with L $\sim$ 10$^{41.5}$ erg s$^{-1}$ in SDSS post-merger galaxies relative to noninteracting galaxies at $0.01 < z < 0.2$. However, they did find an enhancement of mid-IR AGN for the same sample with an AGN excess $\sim 17$ when compared to a noninteracting control sample.

X-ray studies at higher redshifts provide conflicting results. \cite{2011ApJ...743....2S} have found an excess of X-ray AGN with L $>$ 10$^{42}$ erg s$^{-1}$ in kinematic close pairs compared to isolated galaxies at $0.25 < z < 1$. A similar X-ray AGN excess in similar luminosity range (L $\ge$ 10$^{42}$ erg s$^{-1}$) was also found in late-stage mergers relative to isolated galaxies by \cite{2014AJ....148..137L} at $0.25<z<1$.
However, \cite{2020ApJ...904..107S} found a lack of enhancement of X-ray AGN with L $\ge$ 10$^{42}$ erg s$^{-1}$ in galaxy pairs at $0.5 < z < 3.0$ from the Cosmological Evolution Survey and the Cosmic Assembly Near-infrared Deep Extragalactic Legacy Survey.

An alternate approach is to look at the merger frequency in AGN host galaxies compared to non-AGN host galaxies. 
\cite{2019MNRAS.487.2491E} have found an interacting fraction in both optical-selected and mid-IR-selected AGN host galaxies, which is a factor of 2 higher than that in non-AGN control galaxies. Their results strongly suggest that AGN are related to galaxy mergers. Similar results were also found in studies of optical AGN host galaxies at $0.1<z<0.4$ \citep{2008ApJ...677..846B, 2010MNRAS.403.2088L, 2012MNRAS.426..276B, 2015ApJ...804...34H} and mid-IR AGN at $0.4<z<1$ \citep{2008ApJ...674...80U}, suggesting a merger–AGN connection. 

Studies of the merger rate in X-ray-selected AGN host galaxies provide conflicting results. \cite{2010ApJ...716L.125K} have found a high merger rate (18\% vs 1\%) in the Swift Burst Alert Telescope AGN sample (L $\ge$ 10$^{42}$ erg s$^{-1}$) compared to non-AGN host galaxies at $z < 0.05$, suggesting mergers are the dominant mechanism triggering luminous AGN.
Studies of X-ray-selected and optical-selected luminous AGN (L $\ge$ 10$^{42}$ erg s$^{-1}$) also found a higher merger rate compared to non-AGN host galaxies at $z<0.2$ \citep{2012ApJ...746L..22K, 2013MNRAS.431.2661C}.
On the contrary, \cite{2009ApJ...691..705G} analyzed the morphology of X-ray-selected AGN host galaxies with L $\ge$ 10$^{42}$ erg s$^{-1}$ at $0.3 < z < 1.0$ and found no difference between the asymmetry of AGN host galaxies and non-AGN control galaxies.
Numerous similar studies of the morphology of X-ray luminous AGN host galaxies (L $\ge$ 10$^{42}$ erg s$^{-1}$) have not found strong levels of merger signatures at $z<1$ \citep{2005ApJ...627L..97G, 2011ApJ...726...57C, 2013ASPC..477..173C, 2014MNRAS.439.3342V, 2019MNRAS.487.2491E} and at higher redshifts $z\sim2$ \citep{2012ApJ...744..148K}, which suggest that mergers are not the primary mechanism triggering AGN.

\begin{figure*}[t!]
\begin{center}
\begin{minipage}{0.9\textwidth}
\includegraphics[width=\linewidth]{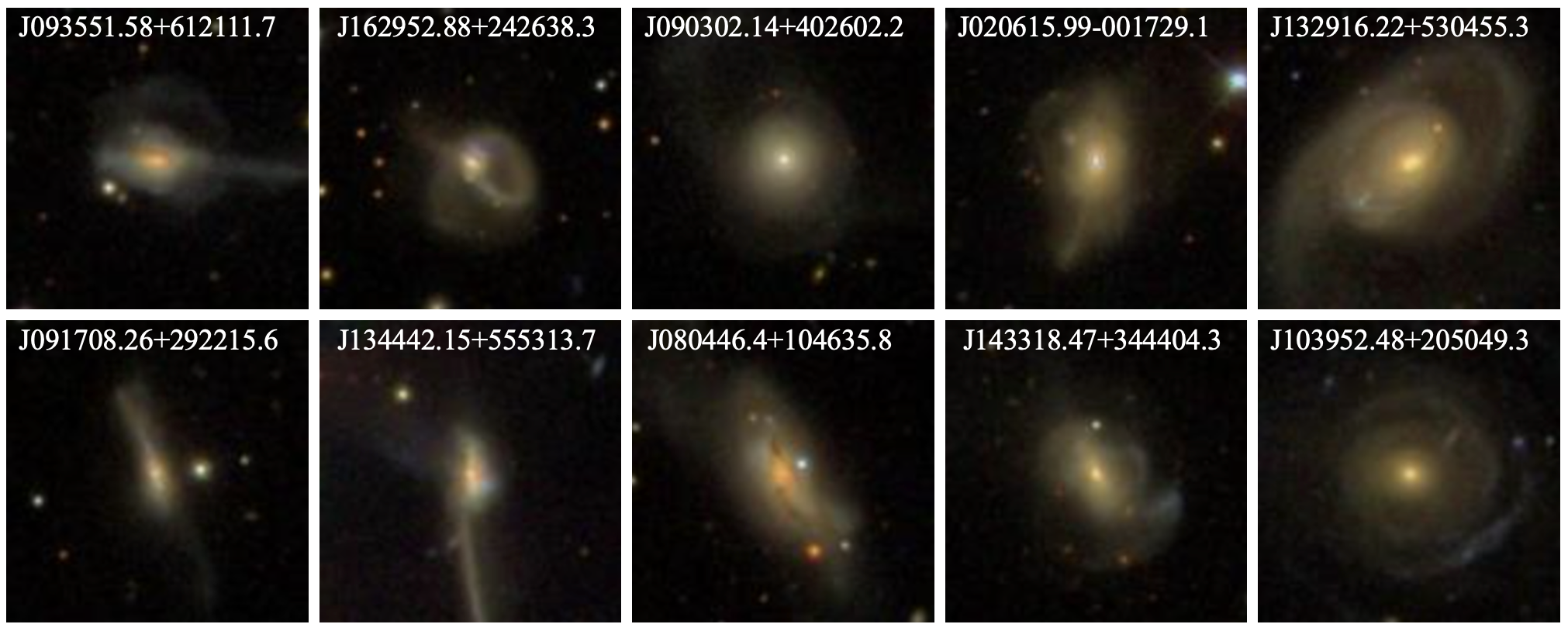}
\end{minipage}
\caption{SDSS montage of 10 random post-mergers with tidal tails, shells, or debris structures. The size of each stamp is $\sim 50''$ on a side. }
\label{fig:PMmontage}
\end{center}
\end{figure*}

\begin{figure*}[t!]
\begin{center}
\begin{minipage}{0.9\textwidth}
\includegraphics[width=\linewidth]{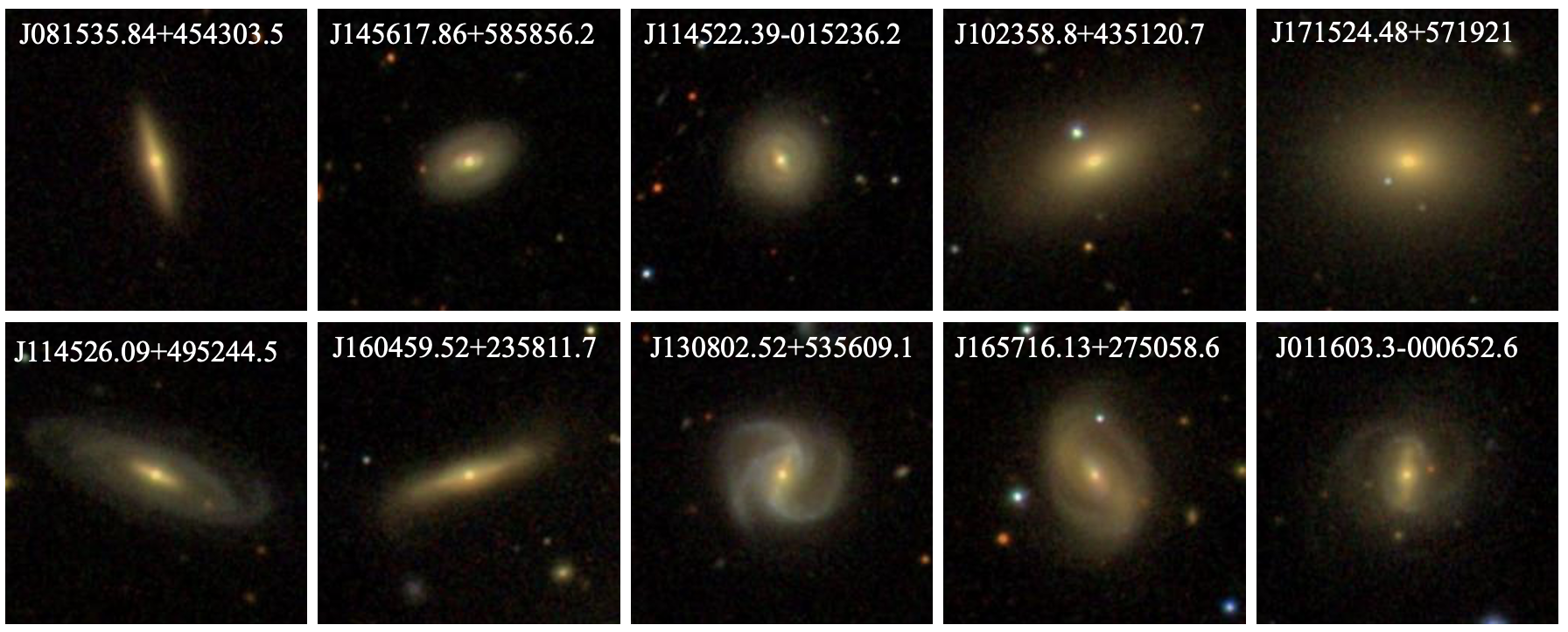}
\end{minipage}
\caption{SDSS montage of 10 noninteracting control galaxies. The control galaxies have a wider range of morphology. Each control galaxy corresponds to one post-mergers in the same order as in Figure~1. The size of each stamp is $\sim 50''$ on a side. }
\label{fig:Controlmontage}
\end{center}
\end{figure*}

There are many potential causes for these discrepant results. According to simulations, AGN luminosities and lifetimes are variable at different merger stages. 
AGN in mergers are in a lower-luminosity phase ($L_{bol}<10^{42}$ erg s$^{-1}$) for a longer duration ($\sim 1-2$ Gyr, \citealp{2008ApJS..175..356H, 2018MNRAS.478.3056B}) compared to the luminous AGN phase ($L_{bol}>10^{42}$ erg s$^{-1}$) of $\sim 1-10$ Myr \citep{2006ApJ...638..686C,2004cbhg.symp..169M, 2009ApJ...698.1550H}. Therefore, studies that do not probe low-luminosity AGN (LLAGN) cannot provide a complete census of the merger-triggered AGN population. The sporadic nature of the luminous AGN phase will also naturally lead to an underestimate in the fraction of mergers that have experienced an AGN. Most previous studies have mainly focused on the luminous AGN phase. Many have relied on single-wave-band AGN diagnostics (such as optical, X-ray, or IR) which provide an incomplete census of AGN. For example, optical AGN diagnostics can suffer from significant contamination. As galaxy mergers can trigger both star formation and AGN, the emission from faint AGN can be overpowered by star formation. In addition, the source of Low-Ionization Nuclear Emission-line Regions (LINERs) and composite emission is also uncertain. X-ray AGN emission can also be contaminated by emission from X-ray binaries or hot gas. Dust obscuration can also lead to an underestimate of the AGN fraction. Lastly, many previous works have relied on incomplete/impure samples of mergers at a range of merger stages, which can obfuscate correlations between mergers and AGN. 

To probe the merger–AGN connection, we need to identify LLAGN in a merger stage where the lifetimes of the merger signatures are comparable to the LLAGN lifetimes. Post-merger galaxies are the ideal sample to investigate this correlation. They are the end products of major mergers observed during or after coalescence, from just before the ultraluminous infrared galaxy (ULIRG) phase through to the remnant phase when the AGN fades and shells and tidal tails are apparent (see Figure~\ref{fig:PMmontage}). These tidal structures have lifetimes comparable to the lifetimes of LLAGN ($\sim$ 1 Gyr, \citealp{1992ApJ...393..484B, 2006ApJ...638..686C, 2008MNRAS.391.1137L, 2014A&A...566A..97J}).  Hence, they are the ideal targets to look for LLAGN.

Unfortunately, identifying AGN in post-merger systems using optical diagnostics (i.e., Baldwin–Phillips–Telervich (BPT) diagrams, \cite{1981PASP...93....5B}) alone is challenging. 
While LINER- and composite-like emission ratios can be caused by AGN, they can also be produced by post-AGB stars or shocks \citep{2013A&A...558A..43S, 2017MNRAS.470.4974D}. As post-merger galaxies are shocked systems, they exhibit LINER-like emission line ratios and cannot be confirmed as AGN by optical diagnostics alone. Therefore, multiwavelength diagnostics are required in order to confirm their AGN nature. 

In this paper, we use these different selection techniques to identify AGN in our post-merger galaxies, which should enable a more comprehensive study of the merger–AGN connection. This paper is arranged as follows. In Section 2, we describe our high-mass post-merger subsample, the available multiwavelength data, and the analysis techniques we used for reducing X-ray observations. In Section 3, we describe the results of the multiwavelength AGN fraction in high-mass post-merger galaxies and control galaxies. In Section 4, we compare our results to previous studies and explain the reasons of the discrepancy. In Section 5, we summarize the main conclusions in this work. 

\section{Sample and Analysis Techniques}  \label{sec:sample}

\subsection{High-mass Post-merger Subsample}
The post-merger sample is derived from a parent catalog of  113,693 galaxies with 0.02 $\le$ z $\le$ 0.06 and 9 $\le$ log M$_*$/M$_{\odot} \le$ 12 from the SDSS Data Release 14 \citep{2000AJ....120.1579Y} main sample. Nair (2023, in preparation; hereafter N23) visually classified the SDSS images of this 113k sample following the prescription described in \cite{2010ApJS..186..427N} to identify bars, rings, spiral arms, interacting galaxies, close pairs, dual nuclei, merger remnants, etc.
N23 identifies $\sim$ 1,200 post-merger remnant galaxies. The work of this paper focuses on the high-mass (log M$_*$/M$_{\odot} \ge$ 10.5) subsample of 807 post-merger galaxies. Example images of post-merger galaxies are shown in Figure~\ref{fig:PMmontage}.

To investigate the X-ray AGN nature of post-merger galaxies, we used new, deep, Chandra observations (proposal ID: 19700403 by PI Nair) for a sample of 12 objects. Eleven of these objects are in our high-mass subsample. The observations were designed to be deep enough to detect a source down to  luminosity of $10^{40}$ erg s$^{-1}$, a threshold chosen as the luminosity from contaminants such as X-ray binaries (XRBs), which are expected to be lower than this threshold for our sample (see Section~3). 
To complement data in the archive, the sample was chosen to not be Seyfert galaxies as classified by the [O III]/H$\beta$ vs [N II]/H$\alpha$ optical BPT diagnostic using the \cite{2001ApJ...556..121K} and \cite{2007MNRAS.382.1415S} demarcation lines and to have z$<$0.04 to keep exposure times reasonable per target.

We crossmatched our high-mass (log M$_*$/M$_{\odot} \ge$ 10.5) post-merger sample with the Chandra and XMM-Newton archive. 
Given the large imaging field, we used a search radius of $10'$ in Chandra and $18'$ in XMM-Newton observations to allow a search for both targeted and serendipitous observations for our galaxies. We also restricted our matches to be within $10'$ of the focal point (off-axis angle) of Chandra observations and $18'$ of the XMM-Newton observations to allow robust point-spread function (PSF) corrections.
There are 36 post-mergers that were observed in Chandra, and 47 were observed in XMM.
We also crossmatched the post-merger sample with the CSC 2.0 catalog \citep{2010ApJS..189...37E} and the 4XMM-DR11 source catalog \citep{2020A&A...641A.136W} to ensure no X-ray source is missed. 
For each observation, we calculated the detection limit in photon counts by using the $3\sigma$ Poisson upper limit for (assumed) zero background count \citep{1986ApJ...303..336G}. We converted the detection limit from counts to fluxes and luminosities using the WebPIMMS\footnote{ \url{https://heasarc.gsfc.nasa.gov/cgi-bin/Tools/w3pimms/w3pimms.pl}} by assuming a simple power-law model with $\Gamma=1.9$ and Galactic absorption assuming the HI4PI survey map \citep{2016A&A...594A.116H}.

In order to search for the very faint AGN, we further selected observations that were deep enough to detect a source with L $\ge 10^{40.5}$ erg s$^{-1}$. (Very few of the archived sources had detection limits down to $10^{40}$ erg s$^{-1}$).  There are 26 Chandra-observed sources and 41 XMM-observed sources that satisfy this detection threshold. An additional Chandra post-merger galaxy has shallower Chandra observations but has a point-source detection. Hence, we still included it in our sample.
Of the 27 Chandra-observed post-mergers, 12 were serendipitously observed in the field of view while the remaining 15 were targeted X-ray observations. Of the 15, 10 were targeted because they were AGN preselected in either hard X-ray, radio, IR, or optical. The remaining 5 were targeted for other reasons not directly related to their potential AGN nature. For the 41 XMM-observed sources, 28 were serendipitously observed while the other 13 post-mergers were targeted. Of the 13, 10 were targeted because they were preselected as AGN in other wavebands. This yields a sample of 38 Chandra-observed galaxies and 41 unique XMM-observed galaxies. 
In total, there are 79 high-mass post-merger galaxies with deep observations or a source detection in either Chandra or XMM. 
A summary of the targets is shown in Table 1 for Chandra data and Table 2 for XMM, which also include details on the X-ray observations. The serendipitously observed galaxies are marked with $^a$ following the SDSS object ID. Galaxies with $^{b}$ marked in Table 1 are the (non-Seyfert) post-mergers observed as part of our Chandra proposal.

Optical data for the 79 post-mergers are available from the value-added catalog compiled by the Max Planck Institute for Astrophysics and the John Hopkins University (MPA-JHU, \citealp{2004MNRAS.351.1151B, 2003MNRAS.341...33K, 2004ApJ...613..898T}). 
The MPA-JHU catalog contains optical emission line fluxes, equivalent widths measurements, and spectral indexes from SDSS single-fiber spectra, which are used to identify AGN in this work. This catalog also provides derived stellar masses and star formation rate (SFR) measured within the 3$"$ SDSS fiber (fiber value) and those measured for the entire galaxy (total value).

We crossmatched these 79 X-ray-observed post-mergers with value-added catalogs from infrared and radio. Infrared measurements were obtained by crossmatching with the All-sky Wide-field Infrared Survey Explorer (ALLWISE) catalog \citep{2014yCat.2328....0C}, which contains photometric measurements for over 700,000,000 objects detected in the all-sky imaging of the Wide-field Infrared Survey Explorer (WISE; \citealp{2010AJ....140.1868W}). This catalog provides magnitudes measured with profile-fitting photometry in four different filters centered at 3.4, 4.6, 12, and 22 $\mu$m (denoted as W1, W2, W3, and W4). By using a search radius of $5''$, we found counterparts for all 79 post-mergers.
Radio data were obtained from the Very Large Array (VLA) Faint Images of the Radio Sky at Twenty cm (FIRST) catalog \citep{1994ASPC...61..165B}, which contains radio flux densities for sources detected by the VLA over 10,000 square degrees of the north and south sky in the FIRST \citep{1995ApJ...450..559B}) survey. Using a search radius of $30''$, 77 out of these 79 post-mergers had counterparts in the VLA FIRST catalog. The remaining two galaxies do not have FIRST radio observation and are excluded from the multiwavelength AGN analysis.

In summary, out of 79 high-mass X-ray observed post-mergers, 77 have multiwavelength observations in all four wavebands: optical, X-ray, infrared, and radio band. We will discuss AGN identification in Section~3.

\vspace{-2cm}
\begin{deluxetable*}{l|c|c|c|c|c|c|c|c|c|c|c|c}
\tablenum{1}
\tablecaption{Chandra-observed post-merger galaxies}
\tabletypesize{\scriptsize}
\tablewidth{0pt}
\tablehead{
\colhead{\textbf{SDSS ObjID}} & \colhead{\textbf{R.A.}} & \colhead{\textbf{Dec.}} & \colhead{\textbf{z}} & \colhead{\textbf{log$\frac{M_*}{M_{\odot}}$}} 
& \colhead{\textbf{ObsID}} & \colhead{\textbf{Exp.}} & \colhead{\textbf{Sep.}} & \colhead{\textbf{$\Gamma$}} & \colhead{\textbf{log$\frac{L_{X}}{erg s^{-1}}$}} & \colhead{\textbf{log$\frac{L_{XRB}}{erg s^{-1}}$}} & \colhead{\textbf{log$\frac{L_{gas}}{erg s^{-1}}$}} &\colhead{\textbf{BPT}}
\\
 & J2000 & J2000 & & & & ksec & arcmin & & & & & }
\decimalcolnumbers
\startdata
1237651272966275163 & 143.96 & 61.35 & 0.039 & 10.9 & 2033 & 49.3 & 0.59 & 0.44 & 41.6 & 39.9 & 39.4 & L \\
1237651539262701677$^a$ & 242.84 & 52.45 & 0.029 & 10.6 & 6859 & 14.8 & 2.58 & - & $<$39.9 & 39.0 & 36.6 & N \\
1237651539792953375 & 222.33 & 63.27 & 0.042 & 11.3 & 12797 & 28.7 & 0.29 & 1.97 & 40.6 & 39.4 & 37.4 & S \\
1237652600101142677$^b$ & 319.27 & -6.72 & 0.029 & 11.1 & 20435 & 12.9 & 0.44 & 1.92 & 41.0 & 39.3 & 38.5 & C \\
1237654030867562543$^a$ & 176.39 & 3.23 & 0.020 & 10.8 & 17029 & 4.9 & 9.38 & 1.9 & 40.4 & 40.3 & 39.8 & C \\
1237654400765395076$^a$ & 177.62 & 65.74 & 0.055 & 10.9 & 10355 & 7.8 & 1.75 & 1.9 & 40.9 & 39.4 & 38.5 & S \\
1237654602104569933$^a$ & 159.88 & 5.10 & 0.027 & 10.7 & 15117 & 13.6 & 4.12 & 1.9 & 40.1 & 39.8 & 39.5 & C \\
1237654881271414816 & 209.28 & 5.25 & 0.040 & 10.8 & 14932 & 46.5 & 0.19 & 1.9 & 40.2 & 40.0 & 39.6 & C \\
1237655349431959652 & 246.76 & 43.47 & 0.047 & 11.0 & 10273 & 14.8 & 1.68 & 1.9 & 40.7 & 39.2 & 36.9 & L \\
1237655465922658349 & 215.65 & 59.93 & 0.029 & 10.8 & 14933 & 29.0 & 0.18 & 1.89 & 40.7 & 39.1 & 38.2 & C \\
1237655468597575708$^a$ & 225.33 & 1.63 & 0.035 & 11.1 & 5907 & 48.4 & 5.76 & 1.01 & 41.7 & 40.4 & 38.5 & C \\
1237656496713105500$^a$ & 333.63 & 13.77 & 0.024 & 10.6 & 5635 & 27.0 & 3.52 & - & $<$39.5 & 38.7 & 36.7 & N \\
1237657629513809952$^b$ & 135.75 & 40.43 & 0.029 & 10.6 & 20430 & 11.9 & 0.43 & - & $<$40.0 & 39.2 & 38.7 & C \\
1237658205567320143 & 130.93 & 35.82 & 0.054 & 11.1 & 18143 & 22.7 & 0.27 & 1.17 & 43.4 & 39.6 & 39.1 & S \\
1237658423556964512$^a$ & 167.20 & 7.00 & 0.041 & 10.9 & 12749 & 19.3 & 1.33 & - & $<$40.1 & 39.3 & 38.1 & N \\
1237658612519141389$^a$ & 173.70 & 49.07 & 0.033 & 11.4 & 6946 & 47.3 & 1.86 & 1.9 & 39.8 & 39.3 & 36.9 & N \\
1237658800968761369$^b$ & 202.31 & 53.08 & 0.029 & 11.2 & 20436 & 12.9 & 0.28 & - & $<$40.0 & 39.1 & 37.5 & L \\
1237659897784566134$^b$ & 264.68 & 57.23 & 0.030 & 10.8 & 20432 & 12.9 & 0.32 & - & $<$40.0 & 39.0 & 37.4 & N \\
1237661354316726385 & 179.77 & 58.34 & 0.054 & 11.4 & 17116 & 93.8 & 0.33 & 1.16 & 42.0 & 39.4 & 37.3 & L \\
1237661387602853939 & 206.17 & 55.88 & 0.037 & 10.9 & 18177 & 59.9 & 0.13 & 1.17 & 41.7 & 39.8 & 39.4 & S \\
1237662194528026666$^b$ & 189.59 & 42.20 & 0.029 & 10.7 & 20431 & 11.9 & 0.26 & 1.64 & 41.4 & 39.0 & 37.0 & L \\
1237662199356325991$^b$ & 225.46 & 9.72 & 0.034 & 10.7 & 20427 & 16.9 & 0.35 & - & $<$40.0 & 39.1 & 36.8 & N \\
1237662224616128524$^a$ & 218.32 & 34.73 & 0.034 & 10.9 & 19668 & 24.7 & 8.06 & 1.78 & 41.7 & 39.9 & 38.1 & L \\
1237662236393406478$^a$ & 186.64 & 9.03 & 0.025 & 11.0 & 2982 & 37.9 & 2.60 & 1.9 & 39.7 & 39.1 & 36.3 & L \\
1237662335719309348$^b$ & 241.66 & 30.09 & 0.022 & 11.0 & 20434 & 7.0 & 0.37 & - & $<$40.0 & 39.6 & 38.6 & C \\
1237662666963025994 & 247.47 & 24.44 & 0.038 & 10.7 & 18169 & 9.1 & 0.18 & 1.52 & 42.3 & 39.7 & 39.3 & C \\
1237663480337072277 & 343.77 & 0.97 & 0.053 & 10.8 & 8173 & 15.9 & 0.29 & 1.9 & 40.8 & 39.3 & 38.7 & N \\
1237663783674314797 & 31.56 & -0.29 & 0.043 & 11.3 & 19560 & 48.6 & 0.36 & 1.57 & 43.0 & 39.8 & 39.3 & L \\
1237665331471515679 & 208.07 & 31.44 & 0.045 & 11.2 & 12712 & 67.8 & 0.28 & 1.07 & 42.9 & 39.7 & 39.0 & L \\
1237665372796223576$^a$ & 239.14 & 20.05 & 0.032 & 11.1 & 21400 & 17.8 & 5.77 & - & $<$40.0 & 39.2 & 36.7 & N \\
1237665532798763287$^b$ & 244.66 & 15.55 & 0.030 & 11.3 & 20437 & 12.9 & 0.41 & - & $<$40.0 & 39.3 & 37.2 & L \\
1237665564998762503$^b$ & 219.84 & 19.99 & 0.030 & 11.0 & 20433 & 12.9 & 0.34 & - & $<$40.0 & 39.0 & 37.7 & C \\
1237665566615666691 & 234.72 & 17.02 & 0.030 & 11.0 & 18080 & 9.8 & 0.16 & 1.01 & 41.4 & 39.1 & 37.9 & S \\
1237667212121604127$^b$ & 138.57 & 22.27 & 0.027 & 10.6 & 20429 & 11.4 & 0.30 & - & $<$40.0 & 38.7 & 37.1 & N \\
1237667430091194665 & 134.65 & 18.37 & 0.060 & 11.1 & 14970 & 19.8 & 0.16 & 1.9 & 40.7 & 39.7 & 39.0 & S \\
1237667430630228069$^b$ & 139.56 & 20.36 & 0.031 & 10.5 & 20428 & 13.8 & 0.42 & 1.9 & 40.4 & 38.9 & 37.5 & L \\
1237667536920707358 & 121.19 & 10.77 & 0.035 & 10.8 & 22089 & 10.0 & 0.28 & 0.43 & 43.2 & 39.6 & 38.8 & S \\
1237667915416731720$^a$ & 176.22 & 19.77 & 0.027 & 10.7 & 17201 & 61.3 & 3.94 & 1.9 & 41.2 & 39.0 & 38.3 & C \\
\enddata
\tablecomments{(1) SDSS object IDs. IDs marked with $^a$ were serendipitously observed in the field of view. IDs marked with $^{b}$ were the objects observed as part of our Chandra proposal (ID: 19700403). 
Columns (2) and (3) are the RA and DEC J2000 epoch coordinates, respectively. (4) Spectroscopic redshifts. (5) Stellar mass in log M$_*$/M$_{\odot}$. (6) Chandra observation IDs. (7) Exposure time in ksec. (8) Off-axis angle in arcminutes. (9) Photon index $\Gamma$. For  sources with counts $>50$, the spectra were fit with XSPEC and the best-fit $\Gamma$ is reported. For sources with counts $<50$, a $\Gamma = 1.9$ is used, which is typical for low-luminosity AGN. See Section 2.3 for details. (10) 2 - 10 keV luminosity calculated by XSPEC or WebPIMMS. Upper limits are shown for nondetections. (11) The expected 2 - 10 keV luminosity from low-mass and high-mass X-ray binaries calculated by using the \cite{2010ApJ...724..559L} empirical relation. (12) The expected 0.5 - 2 keV luminosity from hot gas calculated using the \cite{2012MNRAS.426.1870M} relation. (13) Optical AGN BPT classification: S for Seyferts. L for LINERs. C for composites. N for objects not classified on the BPT due to a low S/N. There are no high-mass post-mergers classified as star forming by the BPT diagnostic.} 
\end{deluxetable*}

\begin{deluxetable*}{l|c|c|c|c|c|c|c|c|c|c|c}
\tablenum{2}
\tablecaption{XMM-observed post-merger galaxies}
\tabletypesize{\scriptsize}
\tablewidth{0pt}
\tablehead{
\colhead{\textbf{SDSS ObjID}} & \colhead{\textbf{R.A.}} & \colhead{\textbf{Dec.}} & \colhead{\textbf{z}} & \colhead{\textbf{log$\frac{M_*}{M_{\odot}}$}} 
& \colhead{\textbf{ObsID}} & \colhead{\textbf{Exp.}} & \colhead{\textbf{Sep.}} & \colhead{\textbf{log$\frac{L_{X}}{erg s^{-1}}$}} & \colhead{\textbf{log$\frac{L_{XRB}}{erg s^{-1}}$}} & \colhead{\textbf{log$\frac{L_{gas}}{erg s^{-1}}$}} &\colhead{\textbf{BPT}}
\\
 & J2000 & J2000 & &  &  & ksec & arcmin & & & & }
\decimalcolnumbers
\startdata
1237649963529535543 & 51.36 & -6.14 & 0.034 & 10.9 & 0760230401 & 21.9 & 1.70 & 42.4 & 40.3 & 39.9 & S \\
1237651534946959533 & 130.40 & 54.91 & 0.045 & 10.7 & 0403190901 & 13.5 & 1.68 & 42.7 & 39.7 & 38.8 & S \\
1237651735777509473$^a$ & 226.99 & 1.23 & 0.035 & 10.9 & 0402781001 & 19.7 & 4.28 & 40.9 & 40.3 & 39.9 & C \\
1237652900758880309 & 24.98 & -9.24 & 0.042 & 11.6 & - & - & - & 40.5 & 40.6 & 38.0 & L \\
1237652901297061972$^a$ & 27.88 & -8.78 & 0.036 & 11.2 & 0761730401 & 37.3 & 9.64 & $<$39.4 & 40.2 & 39.3 & N \\
1237652942625767592 & 330.42 & 11.85 & 0.030 & 11.3 & 0762630201 & 21.9 & 1.75 & 40.3 & 40.3 & 39.0 & L \\
1237653438155128872 & 346.18 & -8.68 & 0.047 & 11.4 & 0790640101 & 53.6 & 1.71 & 44.1 & 40.6 & 40.0 & S \\
1237653616933535826$^a$ & 158.19 & 58.86 & 0.046 & 11.5 & 0142970101 & 25.5 & 9.54 & $<$39.8 & 40.5 & 37.8 & L \\
1237654382515060810$^a$ & 138.35 & 52.98 & 0.025 & 11.1 & - & - & - & 40.6 & 40.1 & 37.6 & L \\
1237655370892247130$^a$ & 154.65 & 59.66 & 0.044 & 10.7 & 0406630201 & 41.4 & 11.22 & $<$39.1 & 40.0 & 39.5 & C \\
1237655373573062713$^a$ & 247.43 & 40.81 & 0.029 & 11.6 & 0203710201 & 26.3 & 1.07 & 41.6 & 40.5 & 38.0 & L \\
1237657857145962520 & 176.84 & 52.44 & 0.049 & 10.7 & 0504101401 & 30.3 & 1.71 & 41.5 & 40.2 & 39.8 & S \\
1237657857145962550$^a$ & 176.92 & 52.44 & 0.048 & 10.8 & 0504101401 & 30.3 & 4.01 & 41.0 & 39.9 & 39.0 & C \\
1237658297922945185$^a$ & 157.49 & 5.48 & 0.056 & 10.6 & 0148560501 & 103.8 & 9.09 & $<$38.9 & 39.6 & 37.4 & N \\
1237658299534802987 & 160.29 & 6.82 & 0.033 & 10.6 & 0782530701 & 9.9 & 1.69 & 40.4 & 39.6 & 38.6 & C \\
1237658304352944175$^a$ & 160.60 & 58.45 & 0.045 & 10.6 & - & - & - & 40.7 & 39.6 & 37.5 & L \\
1237658304355172477$^a$ & 170.13 & 59.87 & 0.054 & 10.9 & 0502780201 & 21.2 & 14.70 & $<$39.5 & 39.9 & 37.4 & N \\
1237660613973049348 & 168.45 & 9.58 & 0.029 & 10.9 & 0601780201 & 15.1 & 1.68 & 42.8 & 40.1 & 39.6 & S \\
1237661416605679815$^a$ & 224.94 & 49.50 & 0.027 & 10.6 & 0401270301 & 23.1 & 3.95 & $<$38.9 & 39.7 & 38.9 & N \\
1237661949739663472 & 235.86 & 7.91 & 0.036 & 10.8 & 0822391101 & 16.2 & 1.69 & 42.2 & 40.1 & 39.6 & C \\
1237662236394258511$^a$ & 188.68 & 9.00 & 0.043 & 10.6 & 0722960101 & 52.9 & 6.96 & 40.7 & 40.2 & 39.9 & C \\
1237662340012507431$^a$ & 239.43 & 27.46 & 0.032 & 10.8 & - & - & - & 40.5 & 39.8 & 38.9 & C \\
1237664092903047228$^a$ & 138.74 & 29.73 & 0.021 & 10.9 & 0673150401 & 42.4 & 13.55 & 40.7 & 39.9 & 38.8 & C \\
1237664106852319509$^a$ & 149.77 & 13.02 & 0.037 & 10.7 & 0504100201 & 21.9 & 1.68 & 41.6 & 39.8 & 38.8 & S \\
1237664106852450313$^a$ & 149.98 & 13.04 & 0.035 & 10.8 & 0504100201 & 21.9 & 10.56 & 43.3 & 40.8 & 40.5 & C \\
1237664667902345246$^a$ & 166.34 & 38.02 & 0.030 & 10.7 & - & - & - & 40.0 & 39.7 & 37.3 & L \\
1237664880485728337$^a$ & 139.28 & 29.37 & 0.035 & 10.7 & 0673150301 & 31.9 & 9.32 & 41.2 & 40.0 & 39.5 & S \\
1237665025440677967$^a$ & 181.45 & 35.17 & 0.054 & 10.6 & - & - & - & 43.6 & 40.1 & 39.7 & C \\
1237665532247670906$^a$ & 211.42 & 25.23 & 0.030 & 10.8 & - & - & - & 42.0 & 39.8 & 38.0 & S \\
1237666299481817317$^a$ & 50.18 & -1.04 & 0.021 & 11.1 & 0810600201 & 43.0 & 1.86 & 40.0 & 40.0 & 37.8 & L \\
1237667107428565094$^a$ & 130.67 & 19.65 & 0.055 & 11.2 & - & - & - & 40.7 & 40.2 & 38.5 & N \\
1237667210529275951 & 184.58 & 29.25 & 0.048 & 10.7 & 0762630301 & 21.9 & 1.70 & 41.0 & 39.7 & 38.4 & S \\
1237667211597119522$^a$ & 169.09 & 29.25 & 0.045 & 11.4 & - & - & - & 41.1 & 40.4 & 38.0 & L \\
1237667255624990807 & 181.18 & 31.17 & 0.025 & 10.8 & 0601780601 & 39.5 & 1.69 & 41.9 & 40.0 & 39.3 & S \\
1237667735572381852 & 159.96 & 20.84 & 0.046 & 11.2 & 0059800101 & 14.9 & 1.66 & 40.9 & 40.1 & 37.8 & L \\
1237668298739351591$^a$ & 181.04 & 20.40 & 0.024 & 11.2 & 0112270601 & 12.4 & 9.71 & 41.2 & 40.2 & 37.6 & L \\
1237671140947591227 & 180.24 & 6.80 & 0.036 & 10.7 & 0312191701 & 12.9 & 1.76 & 42.4 & 39.7 & - & S \\
1237671932283256842$^a$ & 164.93 & 10.07 & 0.035 & 10.9 & 0802200801 & 35.9 & 7.93 & 40.8 & 39.9 & 37.7 & N \\
1237671932820259064$^a$ & 165.31 & 10.48 & 0.042 & 10.9 & - & - & - & 40.3 & 39.9 & 38.6 & N \\
1237673808121495726$^a$ & 128.93 & 21.59 & 0.042 & 10.6 & 0803953001 & 38.8 & 14.55 & $<$39.1 & 39.9 & 39.4 & C \\
1237674462024106294$^a$ & 136.01 & 1.45 & 0.053 & 11.0 & 0725310131 & 12.5 & 11.45 & 41.3 & 40.2 & 39.5 & C \\
\enddata
\tablecomments{(1) SDSS object IDs. Galaxies with $^{a}$ marked are the serendipitous galaxies in XMM observations. (2) and (3) J2000 epoch. (4) Spectroscopic redshifts. (5) Stellar mass in log M$_*$/M$_{\odot}$. (6) XMM observation IDs. If a source is not detected in a single archived image but is detected in the 4XMM-DR11 catalog, then their observation IDs and exposure time are not shown in the table. (7) Exposure time in ksec. (8) Off-axis angle in arcmin. (9) 2 - 10 keV luminosity in log unit estimated by WebPIMMS with a fix $\Gamma = 1.9$. If a source is not detected in a single archived image but is detected in the 4XMM-DR11 catalog, the luminosity reported in the catalog is shown here. (10) 2 - 10 keV luminosity contributed from X-ray binaries in log unit calculated by using the \cite{2010ApJ...724..559L} relation. (11) 0.5 - 2 keV luminosity form hot gas in log unit calculated by using the \cite{2012MNRAS.426.1870M} relation. (12) BPT classification: S for Seyferts. L for LINERs. C for composites. N for not on BPT due to a low S/N. There are no post-mergers classified as star forming by BPT.}
\end{deluxetable*}

\begin{figure}
\begin{center}
\begin{minipage}{0.47\textwidth}
\includegraphics[width=\linewidth]{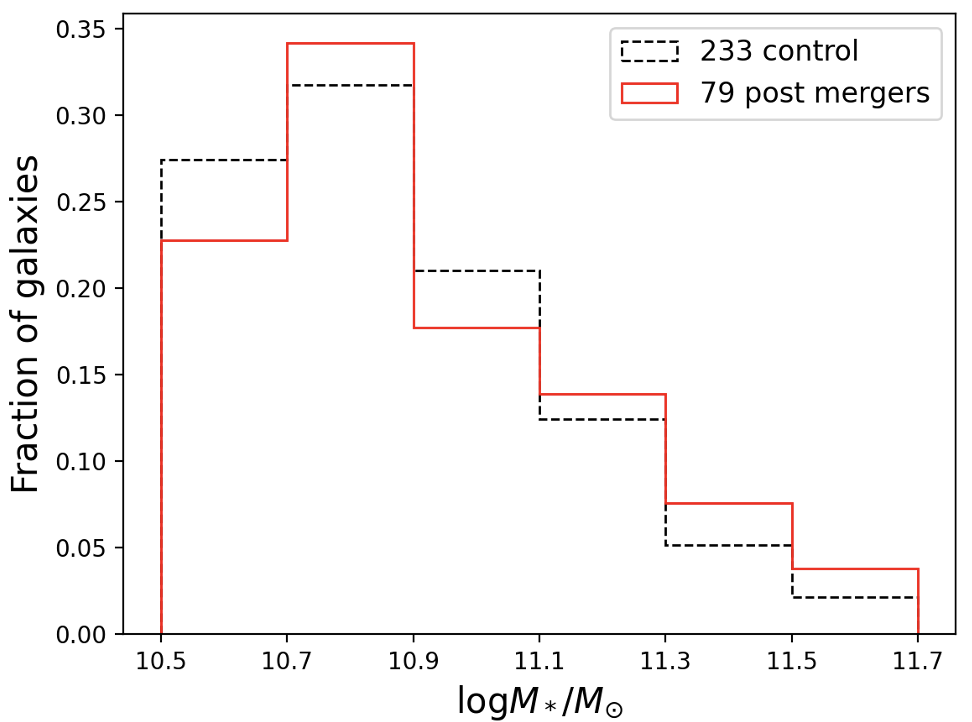}
\includegraphics[width=\linewidth]{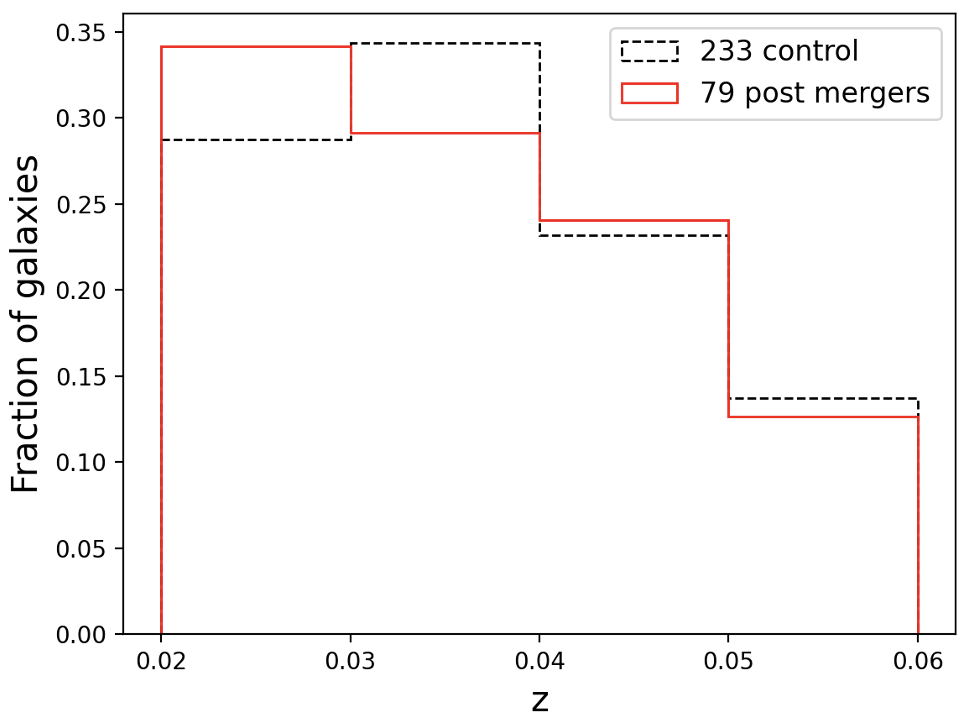} 
\includegraphics[width=\linewidth]{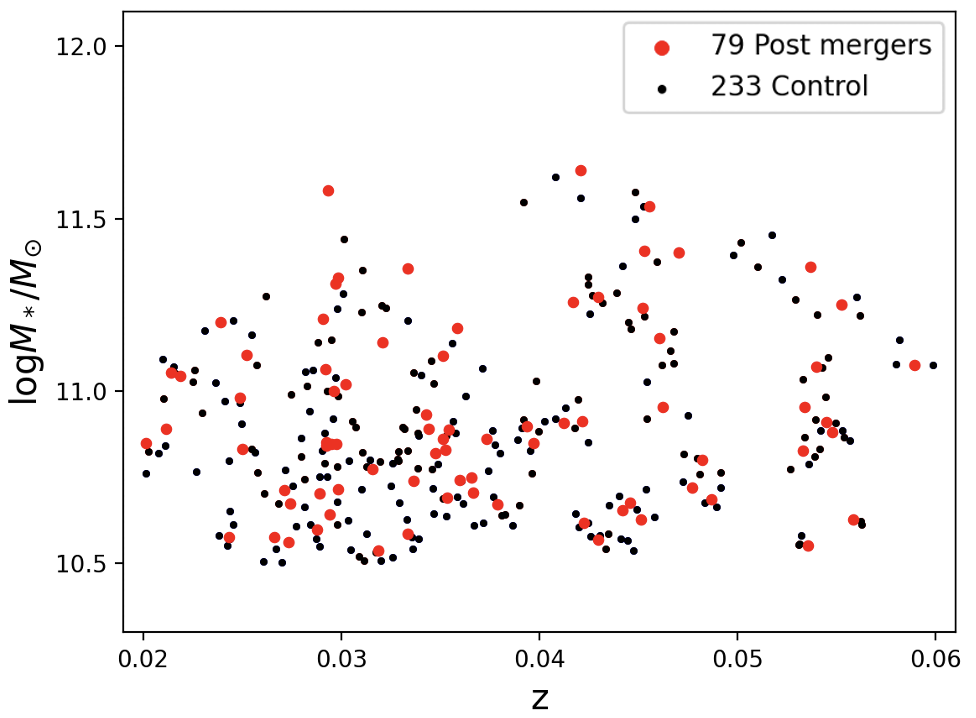} 
\end{minipage}
\caption{The top and middle panels show the stellar mass and redshift normalized distribution of the 79 post-merger galaxies (red) and the 233 control galaxies (black dashed). The bottom panel shows the stellar mass and redshift distribution in the 2D parameter space of post-mergers and control galaxies.}
\label{fig:SampleMassz}
\end{center}
\end{figure}

\subsection{Noninteracting Control Sample}
For the 79 X-ray observed post-mergers, we built a control sample by searching for noninteracting galaxies with similar redshift and stellar mass from the $\sim$ 113,000 N23 parent sample. For each of the post-mergers, we required its control galaxies to have a redshift within $\Delta$z $\pm 0.005$ and stellar mass within $\Delta$log M$_*$/M$_{\odot} \pm 0.1$. We found $\sim$ 19,000 noninteracting control galaxies.
We crossmatched this control sample with the Chandra and XMM archive and calculated the detection limit in luminosity with the same procedure as we did for the post-mergers. We identified 245 Chandra-observed control galaxies and 177 XMM-observed control galaxies for a total of 422 unique control galaxies that were deep enough to detect a source with L $\ge 10^{40.5}$ erg s$^{-1}$. We then required the same number of X-ray-observed control galaxies matched for each of the 79 post-mergers within the $\Delta$z and $\Delta$log M$_*$/M$_{\odot}$ range to avoid any biases. We found 77 out of 79 post-mergers that have three unique X-ray-observed noninteracting control galaxies matched. For the remaining two post-mergers, one has only two unique control galaxies and for the other (one of the most massive galaxy in our sample; log M$_*$/M$_{\odot} = 11.6$) we were not able to find any control galaxies with X-ray observations. Including this object in our post-merger sample did not affect our results so we choose to keep it in the sample. 
In total, there are 233 unique noninteracting control galaxies with deep X-ray observations in either Chandra or XMM. Out of the 233 controls, 129 are observed in Chandra and the other 104 are observed in XMM. Out of the 129 Chandra-observed control galaxies, 21 were targeted objects and 108 were serendipitously observed in the field of view. Out of the 104 XMM-observed galaxies, 12 were targeted and 92 were serendipitous. 
These 233 control galaxies also have been observed with WISE and VLA FIRST. Their optical emission line fluxes, equivalent widths, and spectral indexes are provided by the MPA-JHU catalog.
Figure~\ref{fig:Controlmontage} shows example images of control galaxies. Figure~\ref{fig:SampleMassz} shows the normalized histogram distribution of stellar masses and redshifts of post-merger galaxies and control galaxies as well as the distribution in the 2D parameter space.

\subsection{X-ray Data Reduction}
We downloaded the raw data from the Chandra and XMM archives and reduced all X-ray data in a consistent manner using version 4.10 of the CIAO package with version 4.7.8 of the CALDB calibration files, XSPEC version 12.11.1c, and WebPIMMS version 4.11b. 

For Chandra sources, the data-reduction process is as follows. We reproduced the Chandra event-2 file in CIAO by running \textit{chandra\_repro} and used a $90\%$ enclosed counts fraction (ECF) aperture to account for PSF correction by running \textit{psfsize\_srcs} in CIAO. To estimate the background, a nearby region $\sim$ 10 times larger was chosen without any sources. 
For XMM sources, we used \textit{wavdetect} in CIAO to determine the source regions, which also accounts for PSF correction.

To identify X-ray detections in Chandra and XMM, we required a S/N $> 5$ detection threshold for point sources with X-ray photon counts $> 25$. For point sources with X-ray photon counts $< 25$, we used small number statistics and required the source flux to be greater than the $3\sigma$ Poisson upper limit of the background \citep{1986ApJ...303..336G}.

For Chandra detections with X-ray counts $>50$, we extracted their X-ray spectra by using \textit{specextract} and calculated their X-ray luminosities by modeling an absorbed power-law spectrum using XSPEC. 
For Chandra detections with X-ray counts $<50$ and all XMM detections, their X-ray luminosities are calculated by applying a power-law model of a fixed photon index $\Gamma =1.9$ with Galactic absorption using WebPIMMS \citep{2009ApJ...705.1336C, 2010A&A...512A..58I}.

\begin{figure*}[!hbtp]
\begin{center}
\begin{minipage}{0.49\textwidth}
\includegraphics[width=\linewidth]{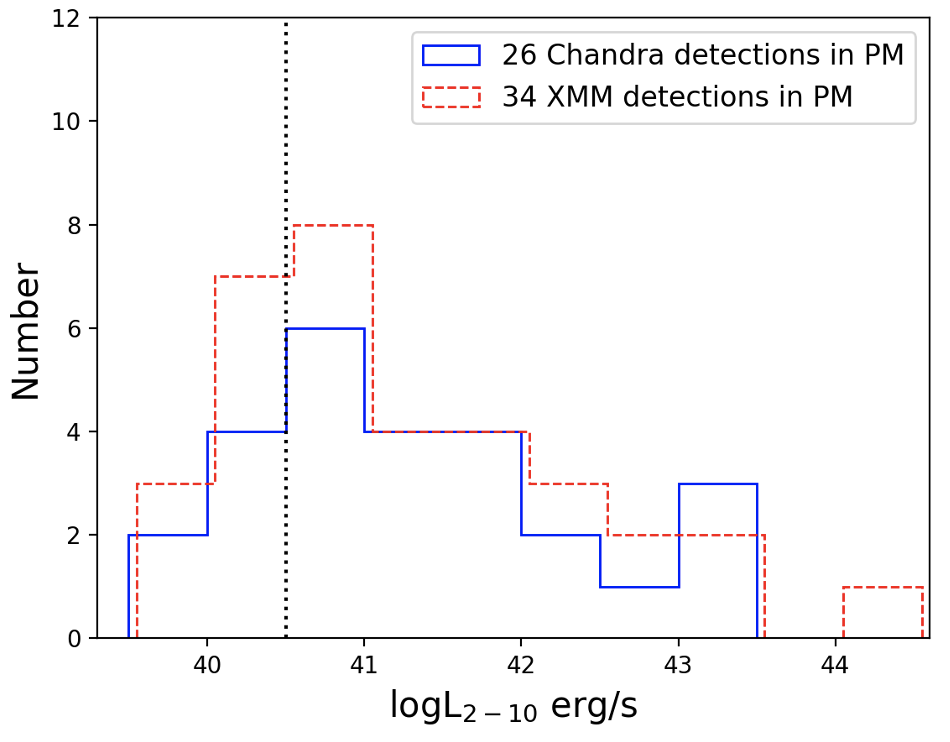}
\end{minipage}
\begin{minipage}{0.49\textwidth}
\includegraphics[width=\linewidth]{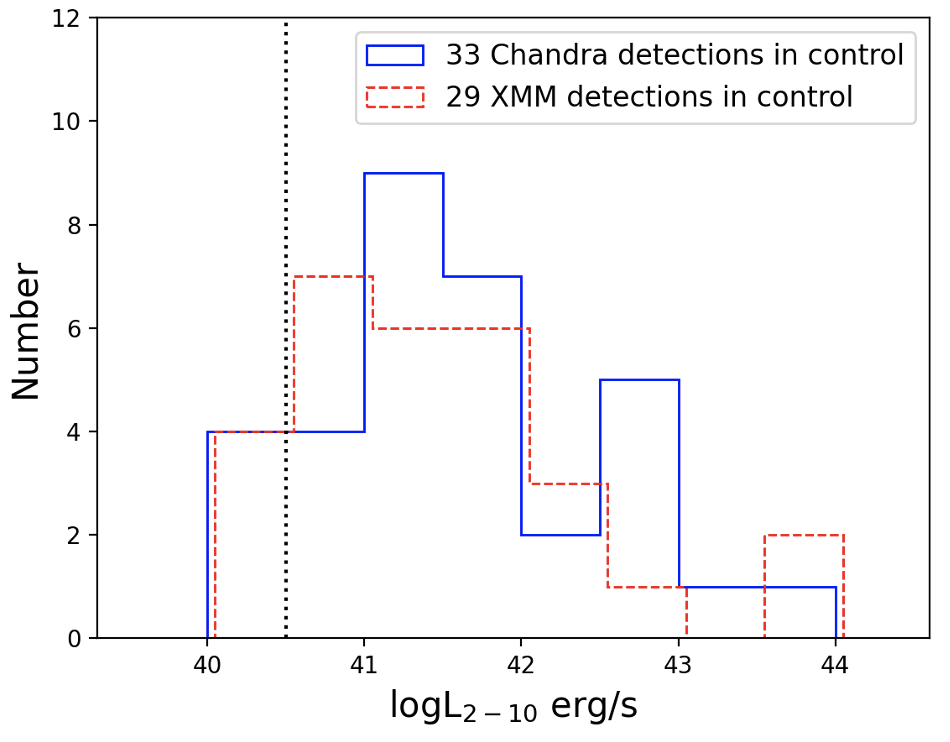}
\end{minipage}
\caption{Distributions of the 2 - 10 keV luminosities of X-ray detections in (left) post-mergers and (right) control galaxies. Chandra detections are shown in blue, and XMM detections are shown in red. The bin size is 0.5 dex. The vertical line is the detection threshold at L = $10^{40.5}$ erg s$^{-1}$. Both the Chandra- and XMM-detected sources span the same range in luminosity, and they also peak at low luminosities around $10^{41}$ erg s$^{-1}$ in both post-mergers and control galaxies.} 
\label{fig:SamplePropertiesXraylum}
\end{center}
\end{figure*}

For the 79 post-mergers, a total of 60 X-ray point sources were detected: 26 X-ray point sources were detected out of 38 (68.4\%) Chandra observations while 34 sources were detected out of 41 (82.9\%) XMM observations. For the 233 control galaxies, a total of 62 X-ray point sources were detected: 33 point sources were detected out of 129 (25.6\%) Chandra observations, while 29 galaxies were detected out of 104 (or 27.9\%) XMM observations. Figure~\ref{fig:SamplePropertiesXraylum} shows the luminosity distribution of X-ray detections in (left) post-mergers and (right) control galaxies. The Chandra detections are shown in blue while the XMM detections are shown in red.  It should be noted that the Chandra observations are well resolved (1$"$), and hence the luminosity corresponds to the central, nuclear luminosity of the galaxy. The XMM observations are not as well resolved, and the luminosity plotted in Figure~\ref{fig:SamplePropertiesXraylum} is for the entire galaxy. However, as can be seen in Figure~\ref{fig:SamplePropertiesXraylum}, the luminosity distribution of the XMM detections is consistent with the Chandra detections, which suggests a similar origin. To investigate this further, we will compare the X-ray AGN fractions in the Chandra vs XMM observations separately. In the next section, we determine how many of the X-ray detections are classified as AGN by comparing the observed X-ray luminosities to the luminosities expected from X-ray binaries and/or hot gas. 

\section{Multiwavelength AGN Diagnostics}
\subsection{X-ray AGN Diagnostics} 
We consider X-ray detections with luminosities higher than those contributed from XRBs and hot gas as X-ray AGN. 
X-ray hot gas mostly contributes to soft X-ray emission between 0.5 – 2 keV with negligible contribution beyond 2 keV while XRBs and AGN are the main sources of X-ray emission beyond 2 keV (see \citealp{1989ARA&A..27...87F} and \citealp{2006ARA&A..44..323F} and references therein).
To identify the source of the X-ray emission, we use the following empirical scaling relation from \cite{2010ApJ...724..559L} to determine the expected X-ray contribution from high-mass and low-mass X-ray binaries (HMXB \& LMXB):
\begin{equation}
L_{XRB}=\alpha M_{\ast} + \beta SFR,
\end{equation}
\noindent where $\alpha = (9.05 \pm 0.37) \times 10^{28} $ erg $s^{-1} M_{\odot}^{-1}$ and $\beta=(1.62 \pm 0.22) \times 10^{39}$ erg $s^{-1} (M_{\odot}yr^{-1})^{-1}$. 
The X-ray emission from hot gas is calculated using the relation from \cite{2012MNRAS.426.1870M}: 
\begin{equation}
L_{Gas}=(8.3 \pm 0.1) \times 10^{38} \times SFR
\end{equation}
with a dispersion of 0.34 dex.
For sources detected in resolved Chandra imaging (PSF FWHM $\sim 0.5''$), their X-ray luminosities are calculated within the $90\%$ ECF aperture, with an average radius of $\sim 1''$. Most Chandra-observed sources have a $90\%$ ECF aperture $< 3''$. We use the stellar masses and SFRs within the $3''$ SDSS fiber to estimate the luminosities from X-ray binaries (XRBs) and hot gas in the very center of the galaxies. 
In addition, two Chandra sources in post-mergers and four Chandra sources in the controls have a $90\%$ ECF aperture $> 3''$. For these sources, we use the total stellar masses and SFRs of the entire galaxy to estimate the luminosities from XRBs and hot gas.
For XMM detections, which are unresolved with a PSF FWHM $\sim 6''$, we use the total stellar mass and total SFR to estimate the luminosities from XRBs and hot gas in the entire galaxy.
 
\begin{figure*}[!htbp]
\begin{center}
\begin{minipage}{0.48\textwidth}
\includegraphics[width=\linewidth]{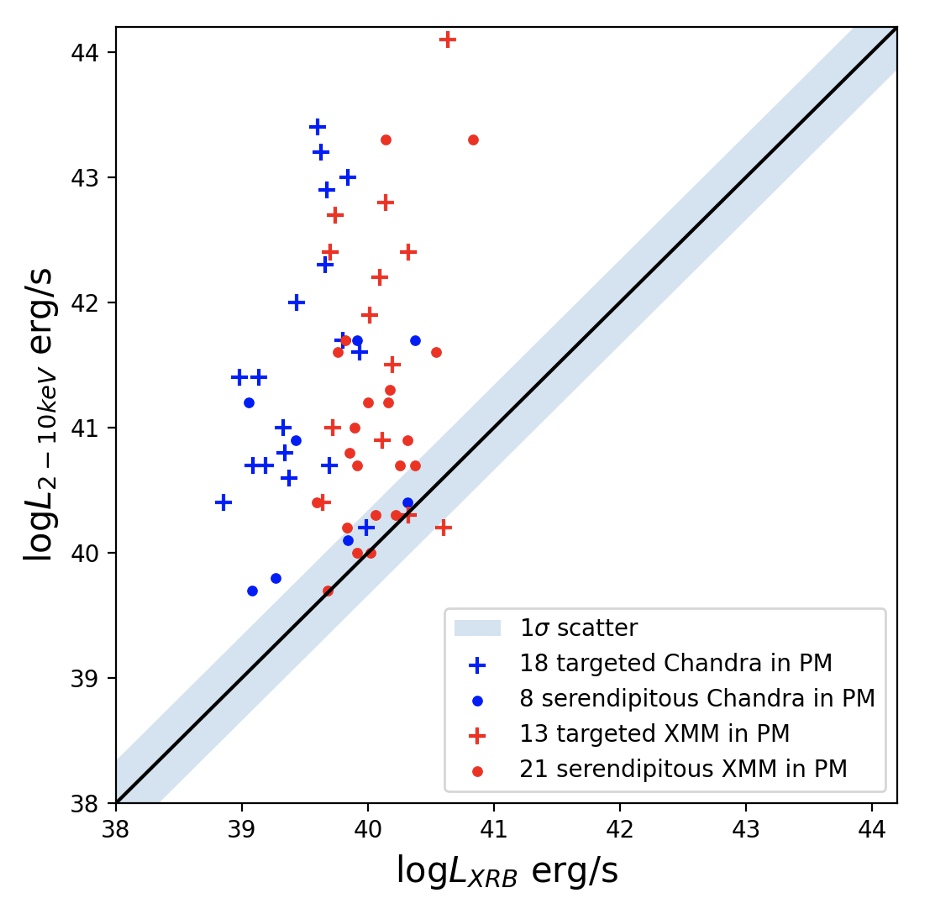}
\end{minipage}
\begin{minipage}{0.48\textwidth}
\includegraphics[width=\linewidth]{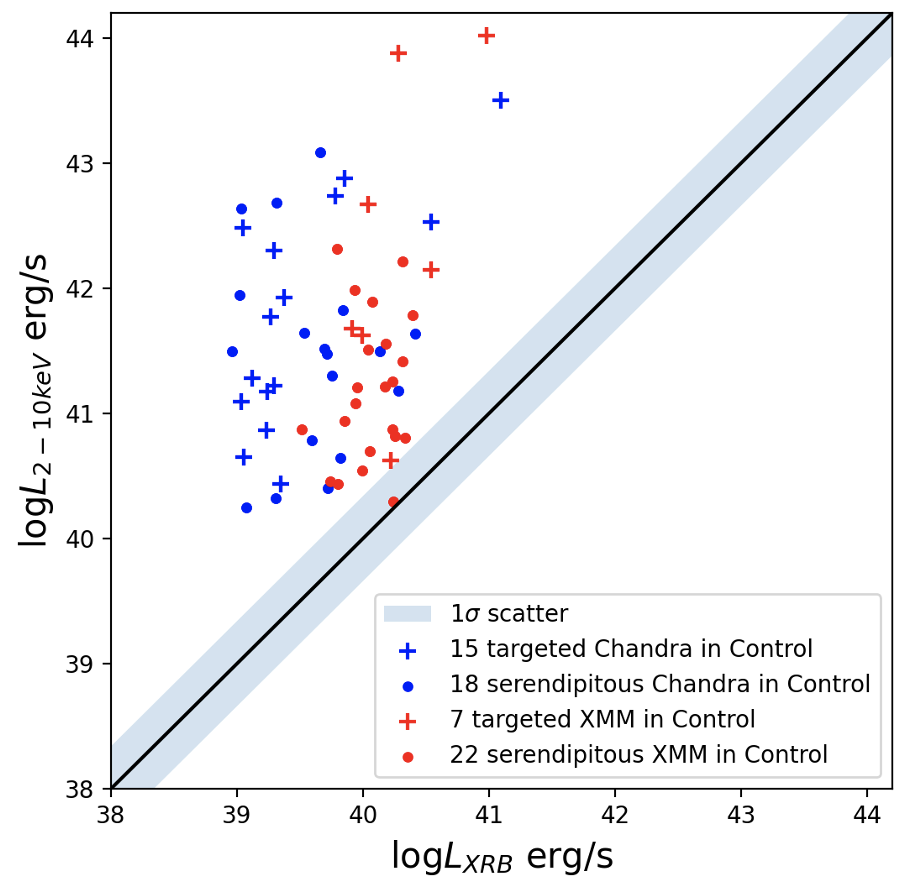}
\end{minipage}
\caption{2 - 10 keV X-ray luminosities vs luminosities estimated from HMXB \& LMXB by the \cite{2010ApJ...724..559L} relation for (left) post-mergers and (right) control galaxies. Chandra detections are shown in blue while XMM detections are shown in red. Targeted sources are shown as crosses and serendipitous sources are shown as circles. The black diagonal line is the $1:1$ relation, with the 1-$\sigma$ scatter of 0.34 dex shown as blue shaded region. We identify detections above the shaded region as X-ray AGN. In post-mergers, 23 AGN are identified in Chandra and 26 AGN are identified in XMM. In control sample, 33 AGN are identified in Chandra, while 28 AGN are identified in XMM.} 
\label{fig:LehmerRelation}
\end{center}
\end{figure*}

\begin{figure*}[!htbp]
\begin{center}
\begin{minipage}{0.48\textwidth}
\includegraphics[width=\linewidth]{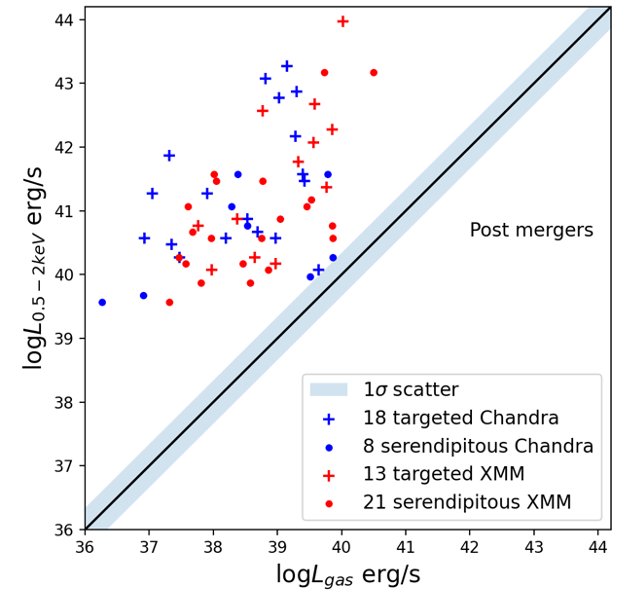}
\end{minipage}
\begin{minipage}{0.48\textwidth}
\includegraphics[width=\linewidth]{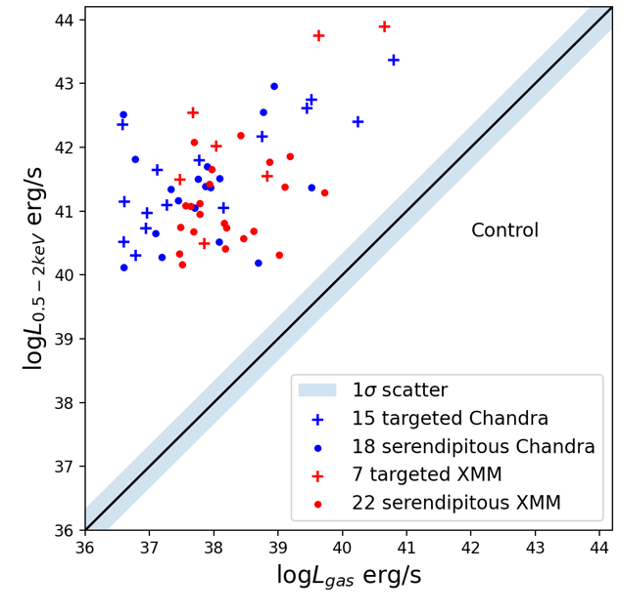}
\end{minipage}
\caption{0.5 - 2 keV X-ray luminosities vs estimated luminosity from X-ray hot gas using the \cite{2012MNRAS.426.1870M} relation for (left) post-mergers and (right) control galaxies. The symbols and legends are the same as in Figure~\ref{fig:LehmerRelation}. The black diagonal line is the $1:1$ relation, with the 1-$\sigma$ scatter of 0.34 dex shown as blue shaded region. All the detections have luminosities (3 orders of magnitudes on average) greater than those estimated from hot gas. The three Chandra post-mergers with luminosities close to the 1:1 line have 2 - 10 keV luminosities consistent with XRBs.}
\label{fig:MineoRelation}
\end{center}
\end{figure*}

Figure~\ref{fig:LehmerRelation} shows the measured 2 - 10 keV X-ray luminosities for our sources vs the luminosities expected from HMXBs and LMXBs using the \cite{2010ApJ...724..559L} relation for (left) post-mergers and (right) control galaxies. The diagonal line shows the $1:1$ relation, and the blue shaded region is the 1-$\sigma$ scatter of the relation. We defined sources above the 1-$\sigma$ scatter of the \cite{2010ApJ...724..559L} relation as X-ray AGN. More or less stringent cuts do not affect our final conclusions but do impact statistics.
Figure~\ref{fig:MineoRelation}, shows the measured 0.5 - 2 keV luminosities of post-mergers (left) and control galaxies (right) vs the estimated luminosity from X-ray hot gas using the scaling relation in \cite{2012MNRAS.426.1870M}. All of our sources have X-ray luminosities higher than that expected from hot gas (on average $\sim$3 orders of magnitude higher). The three Chandra post-merger sources with luminosities close to the 1:1 line also have 2 - 10 keV luminosities consistent with XRBs and are ruled out as AGN.
Out of the 79 X-ray observed post-mergers, 60 are detected, and 49 out of the 60 detections have luminosities higher than those predicted for X-ray binaries (see Figure~\ref{fig:LehmerRelation}) and hot gas (see Figure~\ref{fig:MineoRelation}), which indicates their X-ray emission is dominated by the AGN.
In the control sample of 233 noninteracting galaxies, there are 62 detections, out of which 61 have X-ray luminosities dominated by AGN.

\subsubsection{X-ray AGN Frequency in Post-merger Remnants}
We define the AGN fraction as the number of galaxies hosting an X-ray AGN with L $\ge 10^{40.5}$ erg s$^{-1}$. Any galaxies with a detected X-ray AGN with luminosity $< 10^{40.5}$ erg s$^{-1}$ are not included as not all of our data are deep enough to detect fainter sources. Thus our X-ray AGN fractions are lower limits. 

Out of the 38 Chandra-observed post-mergers, 26 are point-source detections. 
Of the 26 Chandra nuclear detections, 23 have X-ray luminosities higher than that expected from XRBs and hot gas. Hence, they are classified as X-ray AGN. Out of these 23 AGN, 20 have L $\ge 10^{40.5}$ erg s$^{-1}$. This yields an X-ray AGN fraction of $52.6\% \pm 8.1\%$\footnote{The errors are calculated as $\sqrt{f (1-f)/N}$, where f is the AGN fraction and N is the total number of galaxies.} in the Chandra-observed sample. Another three Chandra detections are within the 1-$\sigma$ scatter shaded region and have X-ray luminosities consistent with those of X-ray binaries. Hence, they are not identified as X-ray AGN in our analysis. 

Out of 41 XMM-observed post-mergers, 34 have been detected. Out of the 34 detections, 26 are classified as X-ray AGN, and 24 of the 26 AGN have L $\ge 10^{40.5}$ erg s$^{-1}$. This implies an X-ray AGN fraction of $58.5\% \pm 7.7\%$. The other 8 have luminosities consistent with X-ray binaries and are not classified as X-ray AGN. The XMM AGN fraction in post-mergers is consistent with the Chandra AGN fraction. We take this as confirmation that we can combine the Chandra and XMM samples despite the different imaging resolutions. In total, 44 unique X-ray AGN with L $\ge 10^{40.5}$ erg s$^{-1}$ are identified in the entire X-ray observed post-merger sample of 79 galaxies, with a fraction of $55.7\% \pm 5.6\%$. It should be noted that this is a lower limit on the X-ray AGN fraction. If we include those AGN with detected luminosities $< 10^{40.5}$ erg s$^{-1}$, there will be 49 X-ray AGN identified in 79 post-mergers, with an AGN fraction of $62.0\% \pm 5.5\%$. 

\subsubsection{X-ray AGN Frequency in the Control Sample}
For control galaxies, 
out of the 129 Chandra-observed controls, all 33 X-ray detections 
(blue points in Figure~\ref{fig:LehmerRelation} (right) and Figure~\ref{fig:MineoRelation} (right)) 
are identified as X-ray AGN, and 29 out of the 33 AGN have L $\ge 10^{40.5}$ erg s$^{-1}$. This is an X-ray AGN fraction of $22.5\% \pm 3.7\%$. For the 104 XMM-observed controls, 28 of the 29 X-ray detections are identified as AGN, and 26 out of the 28 AGN have L $\ge 10^{40.5}$ erg s$^{-1}$, which implies an X-ray AGN fraction of $25.0\% \pm 4.2\%$. Again, the XMM AGN fraction is consistent with the Chandra AGN fraction for control galaxies. Combining the two samples, there are 55 unique X-ray AGN with L $\ge 10^{40.5}$ erg s$^{-1}$ identified in the entire X-ray-observed control sample of 233 galaxies. The X-ray AGN fraction in the control sample is $23.6\% \pm 2.8\%$. Including lower-luminosity AGN, there are 61 X-ray AGN in 233 control galaxies, with an AGN fraction of $26.2\% \pm 2.9\%$.

The X-ray AGN frequency in high-mass post-mergers ($55.7\% \pm 5.6\%$) is $\sim$ 2.4 times higher than that in control galaxies ($23.6\% \pm 2.8\%$). This high X-ray AGN excess suggests a strong correlation between galaxy mergers and AGN. The significance of this result can be quantified by a hypergeometric test (urn problem).  A hypergeometric test describes the probability of  `k' successes (the number of AGN) in `n' draws (sample size of post-mergers with Chandra observations) from `N'  objects (total number of post-mergers) containing `K' successes (total number of AGN). In particular:
\begin{equation}
{P(X=k) = {N \choose k} \times {N-K \choose n-k} / {N \choose n}}
\end{equation}
\noindent where ${n}\choose{k}$ = n!/(k! $\times$ (n-k)!) is the binomial coefficient. Assuming a null hypothesis that post-mergers and control galaxies have the same AGN fraction of 23.6\%, the probability that we randomly draw 79 post-mergers from a volume-limited sample of 807 and obtain an AGN fraction of 55.7\% is less than 1 in $10^{10}$. Thus, we can rule out at $>6.5\sigma$ confidence that mergers and control galaxies have similar AGN fractions. Applying a Fisher exact test for the association between post-mergers and X-ray AGN, the probability (p-value) that we observe this excess by chance is only $10^{-7}$. Hence, we find that post-mergers and AGN are strongly correlated.

It should be noted that some of the post-merger and control galaxies were targeted because they were identified as AGN with some other diagnostic. If we restrict our analysis to only the serendipitously observed galaxies, then the X-ray AGN fraction in post-merger galaxies reduces slightly to $45.0\% \pm 7.9\%$, which is consistent with our previous measurement ($55.7\% \pm 5.6\%$). The AGN fraction in the serendipitously observed control galaxy sample also decreases down to $17.0\% \pm 2.7\%$. The AGN excess in post-merger galaxies is then $\sim$ 2.6 times that in control galaxies, consistent with the full sample. Hence we will use the full 79 galaxy post-merger sample and its control going forward.

\subsubsection{Stacking the X-ray Nondetections}
Out of the 38 Chandra-observed post-mergers, 26 have point-source detections and 12 have no individual detections. By stacking the Chandra images of the 12 nondetections, we detected a point source in the center of the stacked image in the soft band (0.5 - 2 keV) and hard band (2 - 8 keV), respectively. 
The source counts are extracted from an aperture of $\sim 3''$ radius after subtracting the average background. 
The average flux is calculated in WebPIMMS by using a power-law spectrum with a fixed photon index $\Gamma =1.9$ and a fixed Galactic absorption column density $n_H =10^{20} cm^{-2}$. The median luminosity distance of the 12 nondetections is used to convert flux to luminosity. 
The stacked source has an average luminosity of log $L= 39.7$ erg s$^{-1}$ with S/N=6 at 0.5 - 2 keV band and log $L= 39.5$ erg s$^{-1}$ with S/N=3 at 2 - 8 keV band.

As hot gas and X-ray binaries can produce X-ray emission in addition to AGN, we need to estimate their contribution to this faint source to determine if the emission is due to AGN, hot gas, or X-ray binaries. 
To estimate the contribution of hot gas, we calculated the count ratio between the 0.5 - 2 keV band and 2 - 8 keV band of the stacked image. For post-mergers, this ratio is $2.5^{+3.3}_{-2.0}$. 
Then we compared it to the expected count ratio of a typical hot gas model assuming an APEC spectrum with 0.768 keV (log T = 6.95) and with the median Galactic absorption column density of the stacked post-mergers in WebPIMMS, which results in a count ratio of $26^{+0.8}_{-0.1}$. The count ratio of the stacked source is much lower than that of the hot gas model. This suggests that hot gas does not contribute much to the X-ray emission of the stacked source. 
We used the median star formation rate of the stacked sample to estimate the X-ray luminosity from hot gas at 0.5 - 2 keV by using the \cite{2012MNRAS.426.1870M} relation (equation 2). We found that hot gas contributes $<3\%$ of the X-ray emission of the stacked source.

To estimate the contribution of XRBs in the 2 - 8 keV band, we applied the \cite{2010ApJ...724..559L} relation with the average stellar mass and SFR of the 12 nondetections. We calculated the X-ray luminosity from XRBs to be log $L= 39.2$ erg s$^{-1}$. The 2 - 8 keV luminosity of the stacked source is within the 1-$\sigma$ scatter of the luminosity from XRBs. Hence, the X-ray emission in the stacked source of post-mergers is more consistent with XRBs rather than AGN or hot gas.

In the stacking analysis of the Chandra-observed control sample, we exclude the observations with targets close to the gaps between the telescope chips or with a bright source nearby. 
By stacking the 83 Chandra nondetections in the control sample, we detected a source of log $L = 39.8$ with S/N=17 at 0.5 - 2 keV and log $L= 39.4$ erg s$^{-1}$ with S/N=8 at 2 - 8 keV. 
The count ratio between the 0.5 - 2 keV band and 2 - 8 keV band of the stacked controls is $3.2 \pm 0.3$. This is much lower than the expected count ratio ($75^{+1.5}_{-1.4}$) of a 0.768 keV hot gas model with the median Galactic absorption column density of the stacked controls. Similarly, we find that hot gas contributes $<1\%$ of the soft-band emission and the X-ray luminosity at the hard band is within the 1-$\sigma$ scatter of the luminosity of XRBs (log $L = 39.7$ erg s$^{-1}$).
Thus, XRBs are the likely source of the X-ray emission in stacked nondetections. However, we still cannot rule out the existence of faint AGN whose luminosity is comparable to XRBs.

We also stacked the XMM nondetections in post-mergers and control galaxies and also found a source in the stacked images in both cases. For both the stacked objects, we found that the X-ray emission is most likely due to XRBs not AGN or hot gas. 

\begin{figure*}[ht!]
\begin{center}
\begin{minipage}{0.49\textwidth}
\includegraphics[width=\linewidth]{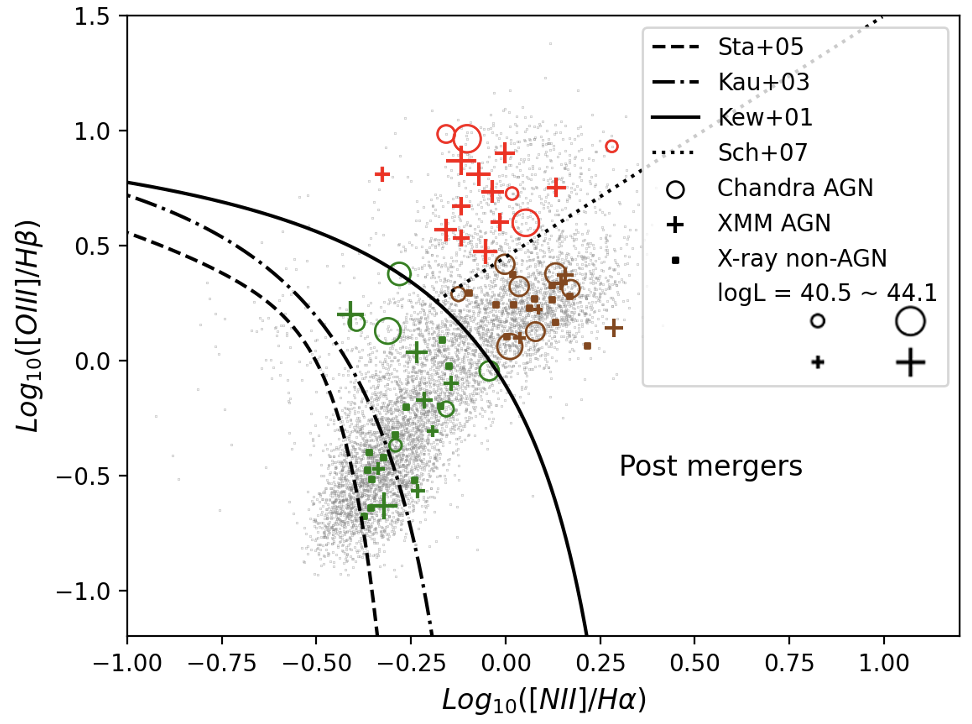}
\end{minipage}
\begin{minipage}{0.49\textwidth}
\includegraphics[width=\linewidth]{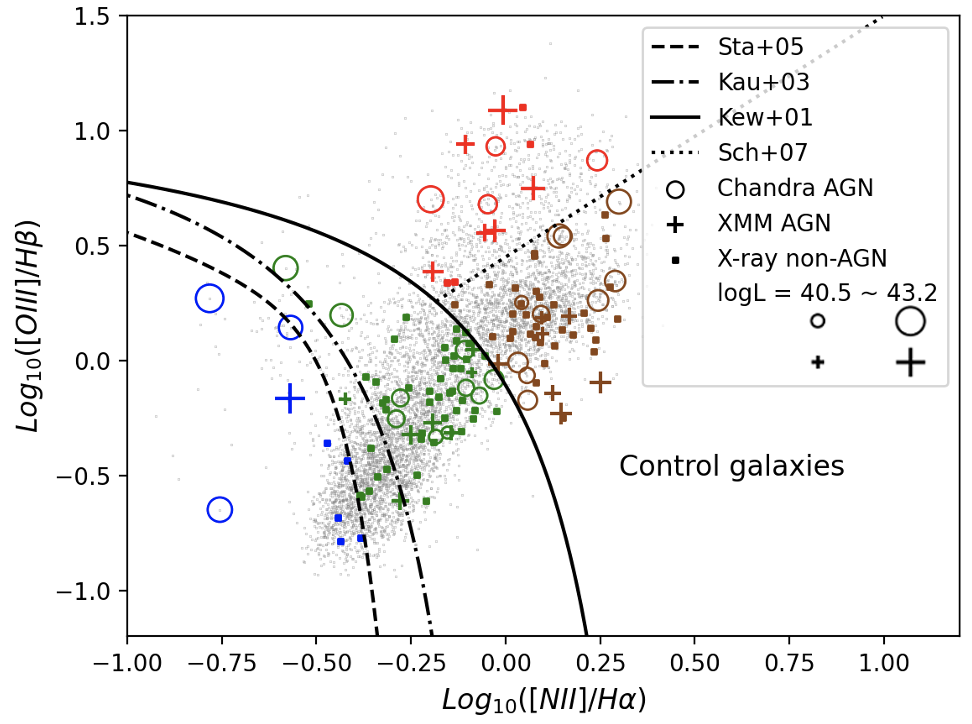}
\end{minipage}
\caption{The [O III]$\lambda5007$/H$\beta$ vs [N II]$\lambda6584$/H$\alpha$ BPT diagram for (left) 79 post-mergers and (right) 233 control galaxies. The grey points are all SDSS  galaxies with log M$_*$/M$_{\odot} \ge 10.5$ drawn from the parent sample. 
Galaxies are classified into Seyferts (red), LINERs (brown), composites (green), and star forming (blue). We consider galaxies above the \cite{2001ApJ...556..121K} solid curve (Seyferts + LINERs) as optical AGN. Seyferts and LINERs are distinguished using the \cite{2007MNRAS.382.1415S} empirical (dotted) line. Objects between the \cite{Stasinska:2006uy} dashed and the \cite{2001ApJ...556..121K} solid lines are composites and those below the \cite{Stasinska:2006uy} line are star forming. The \cite{2003MNRAS.341...33K} empirical curve (dot dashed) is also shown for reference. In 79 post-mergers, there are 16 Seyferts, 24 LINERs, 26 composites, and 13 unclassified. 
All post-mergers including X-ray non-AGN lie above the \cite{Stasinska:2006uy} curve, which indicates that they are AGN candidates. In 233 control galaxies, there are 14 Seyferts, 50 LINERs, 61 composites, 10 star forming and 98 unclassified.}
\label{fig:opticalAGNdiagnostic}
\end{center}
\end{figure*}

\subsection{Optical AGN Diagnostics and Frequency}
To identify optical AGN, we use the emission line ratios [O III]$\lambda5007$/H$\beta$ vs [N II]$\lambda6584$/H$\alpha$ to construct a BPT diagram \citep{1981PASP...93....5B}. 
Fluxes of emission lines are obtained from the MPA-JHU catalog. We place a S/N $\ge$ 3 cut for these four emission lines. By using the \cite{2007MNRAS.382.1415S} empirical relation, the \cite{2001ApJ...556..121K}, and the \cite{Stasinska:2006uy} theoretical relations, galaxies are classified into Seyferts (red, Figure~\ref{fig:opticalAGNdiagnostic}), LINERs (brown), composites (green), and star forming (blue) on the BPT diagram. 
Out of the 79 X-ray observed post-mergers, 66 have S/N $>3$ for all the four lines and are shown in Figure~\ref{fig:opticalAGNdiagnostic}. The remaining 13 galaxies are unclassified due to a low S/N. These 79 galaxies are classified into 16 Seyferts (20.3\%$\pm$4.5\%), 24 LINERs (30.4\%$\pm$5.2\%), 26 composites (32.9\%$\pm$5.3\%), and 13 unclassified (16.5\%$\pm$4.2\%). No galaxy in our high-mass post-merger sample is classified as star forming. (Most galaxies with masses $>10^{10.5} M_{\odot}$ including controls are not classified as star forming.) 
Out of 233 control galaxies, 135 are above the S/N cuts and are classified into 14 Seyferts (6.0\%$\pm$1.6\%), 50 LINERs (21.5\%$\pm$2.7\%), 61 composites (26.2\%$\pm$2.9\%), and 10 star forming (4.3\%)$\pm$1.3\%, as shown in Figure~\ref{fig:opticalAGNdiagnostic} right. The other 98 (42.1\%$\pm$3.2\%) remain unclassified due to a low S/N.

Considering only Seyferts as true AGN, the optical AGN fraction in post-mergers (20.3\%$\pm$4.5\%) is 3.4 times higher than in control galaxies (6.0\%$\pm$1.6\%).
Considering Seyfert+LINERs, the optical AGN fraction in post-mergers (50.6\%$\pm$5.6\%) is 1.8 times higher than that in control galaxies (27.5\%$\pm$2.9\%).
Including composites, the AGN excess in post-mergers relative to controls is 1.6. 
The Seyfert-only excess is more consistent with the X-ray measured AGN excess in post-mergers. Contamination by non-AGN sources (post-AGB stars, shocks) in optically identified LINERs and composites likely causes an underestimate in the measurement of optical AGN excess. This is also true using other optical AGN diagnostics as shown in the Appendix. 

Some studies \citep{2011MNRAS.413.1687C,Stasinska:2006uy} have applied an equivalent width cut on H$\alpha$ (EW$>$3\AA) to remove potential contaminants. 
If we apply the same cut (EW(H$\alpha$)$>$3\AA), we find a Seyfert fraction of 20.3\%$\pm$4.5\% in post-mergers and 5.2\%$\pm$1.4\% in control galaxies. The Seyferts + LINERs fraction is 34.2\%$\pm$5.3\% in post-mergers and 9.0\%$\pm$1.9\% in control galaxies.
We find a consistent optical AGN excess of 3.8 in post-mergers relative to controls when considering either Seyferts or Seyferts+LINERs with the equivalent width cut (See Table 3). However, the equivalent width cut preferentially removes X-ray-identified low-luminosity AGN. 

To investigate this further, we look at the X-ray properties of Seyferts, LINERs, and composites. In Figure~\ref{fig:opticalAGNdiagnostic}, galaxies with an X-ray-identified AGN are shown as circles for Chandra and crosses for XMM. X-ray non-AGN are shown as filled squares. 
Remarkably, all post-merger galaxies classified as Seyferts are also identified as X-ray AGN. 
This high X-ray AGN fraction in Seyferts compared to that in other BPT classes is consistent with a previous study by \cite{2009ApJ...705.1336C}. 
Out of the 24 post-merger LINERs, 11 (or 45.8\% $\pm$ 10.2\%) are identified as X-ray AGN. Out of the 26 composite optical AGN, 14 (or 53.8\% $\pm$ 9.8\%) are identified as X-ray AGN, a similar fraction as LINERs.
Similar trends are seen for controls. Out of the 14 control Seyferts, 10 (or 71.4\% $\pm$ 12.1\%) are identified as X-ray AGN. Out of the 50 control LINERs, 17 (or 34.0\% $\pm$ 6.7\%) are identified as X-ray AGN. Out of 61 composites, 17 (or 27.9\% $\pm$ 5.7\%) are also identified as X-ray AGN. As expected, the X-ray AGN fraction in control galaxies is lower for Seyferts, LINERs, and composites in comparison to post-merger galaxies. While a significant fraction of optical LINERs and composites are clearly X-ray AGN, many are not, suggesting contamination by shocks or post-AGB stars \citep{1994A&A...292...13B,1996ApJS..102..161D,2012ApJ...747...61Y,2013A&A...558A..43S}. Applying an equivalent width cut on H$\alpha$ does not significantly alter these results. We investigate the multiwavelength AGN properties in more detail in Section 4.

\begin{figure*}[t!]
\begin{center}
\begin{minipage}{0.45\textwidth}
\includegraphics[width=\linewidth]{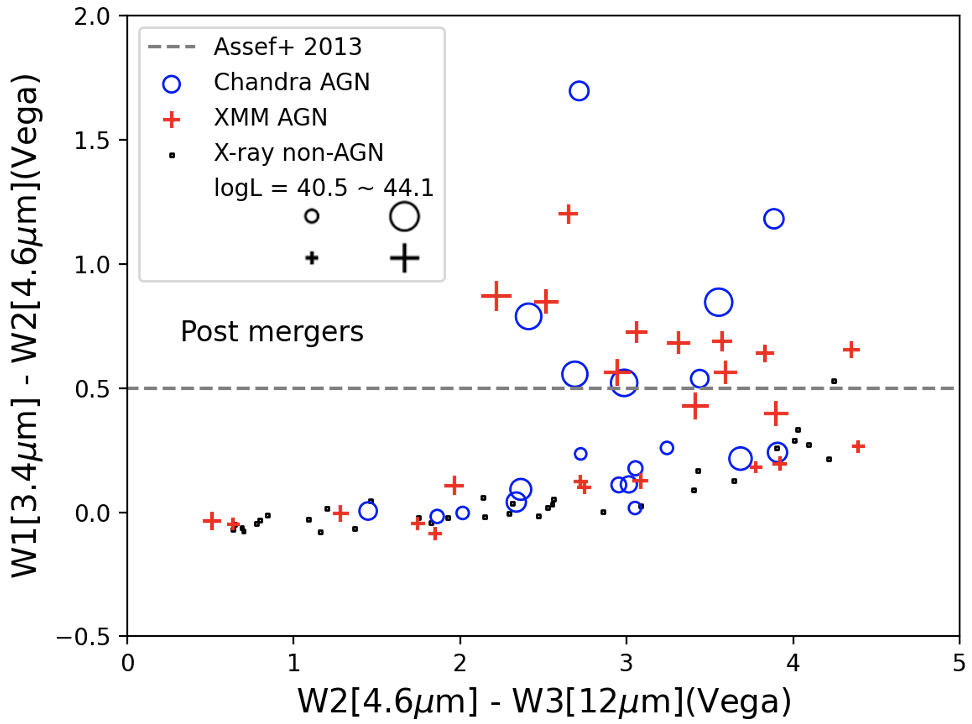}
\end{minipage}
\begin{minipage}{0.45\textwidth}
\includegraphics[width=\linewidth]{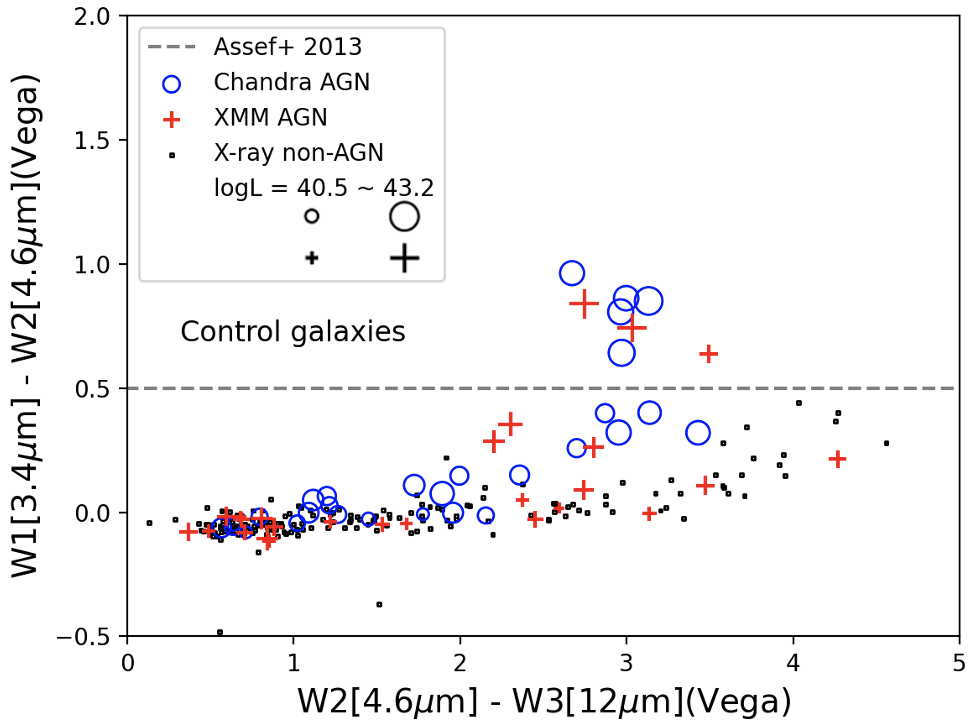}
\end{minipage}
\caption{W1 - W2 vs W2 - W3 in Vega magnitudes for (left) 79 post-mergers and (right) 233 control galaxies. W1, W2, and W3 are the magnitudes of filters centered at 3.4, 4.6, and 12$\mu$m in the WISE survey. The dashed line is the AGN selection criterion adopted from \cite{2013ApJ...772...26A}. Galaxies above the dashed line are classified as WISE infrared AGN. Galaxies with X-ray AGN identified in Chandra are shown as blue circles, while AGN identified in XMM are shown as red crosses. Galaxies not hosting an X-ray AGN are shown as black squares. The sizes of the symbols are related to their X-ray luminosities (for AGN). The relative sizes of the minimal and maximum luminosities are shown in the legend. Out of the 18 WISE AGN identified in post-mergers, 17 are also identified as AGN in X-rays, and they tend to have higher X-ray luminosities than non-WISE AGN. In the control sample, 8 WISE AGN are identified, and all eight are also identified as X-ray AGN. }
\label{fig:wiseAGNdiagnostic}
\end{center}
\end{figure*}

\subsection{WISE AGN Diagnostics and Frequency}
In the infrared, we use the WISE color diagnostics to identify AGN. Figure~\ref{fig:wiseAGNdiagnostic} shows the WISE color W1–W2 vs W2–W3 diagram for the (left) 79 post-mergers and (right) 233 control galaxies. Galaxies with an AGN detected in X-rays are color coded into blue circles for Chandra and red crosses for XMM. The sizes of the data points indicate their X-ray luminosities. By using the WISE AGN criteria: W1 - W2 $>0.5$ from \cite{2013ApJ...772...26A}, we identify 18 WISE AGN out of 79 post-merger galaxies, which implies an infrared AGN fraction of $22.8\% \pm 4.7\%$. Remarkably, 17 out of the 18 WISE AGN are also identified as AGN in X-rays and they are more luminous in X-rays than other X-ray AGN that are not identified as WISE AGN. This is consistent with previous works that show that infrared color diagnostics have a selection bias towards the most luminous AGN \citep{2013MNRAS.434..941M, 2013AJ....145...55Y, 2014MNRAS.438..494R}.

In the control samples, 8 WISE AGN are identified out of the 233 X-ray observed control galaxies, with an infrared AGN fraction of $3.4\% \pm 1.2\%$. All of the 8 AGN are also identified as X-ray AGN. The infrared AGN excess of post-mergers is $\sim$ 6.7 relative to noninteracting control galaxies, which strongly suggests mergers trigger AGN. 
This is consistent with many previous works finding a high infrared AGN excess in post-mergers and galaxy pairs relative to noninteracting galaxies \citep{2013MNRAS.435.3627E, 2014MNRAS.441.1297S, 2017MNRAS.464.3882W, 2018PASJ...70S..37G, 2020A&A...637A..94G}.

\begin{figure*}[t!]
\begin{center}
\begin{minipage}{0.45\textwidth}
\includegraphics[width=\linewidth]{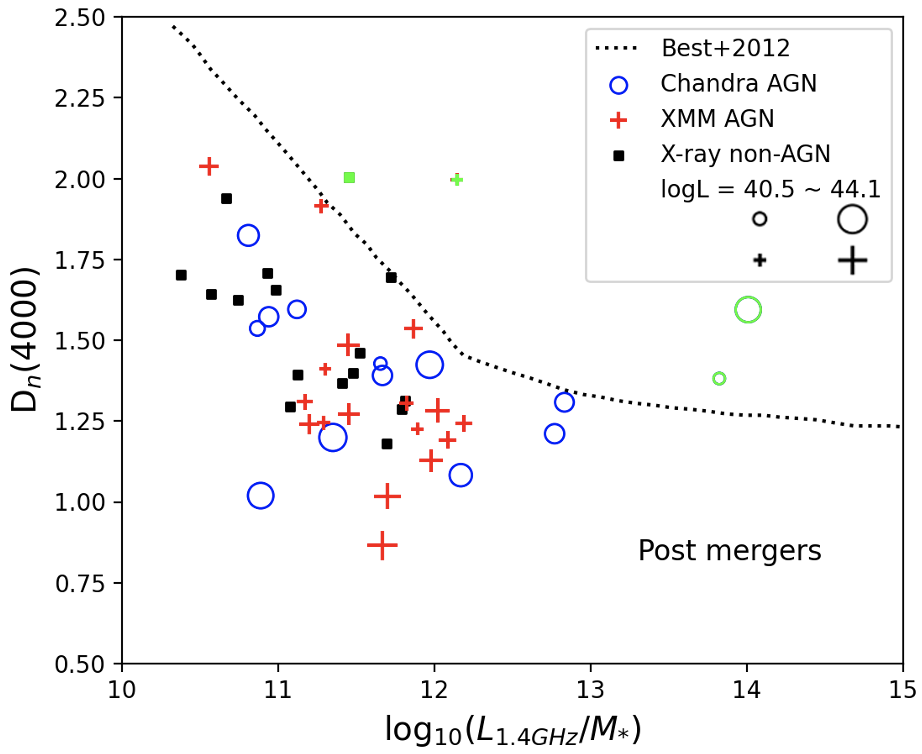}
\end{minipage}
\begin{minipage}{0.45\textwidth}
\includegraphics[width=\linewidth]{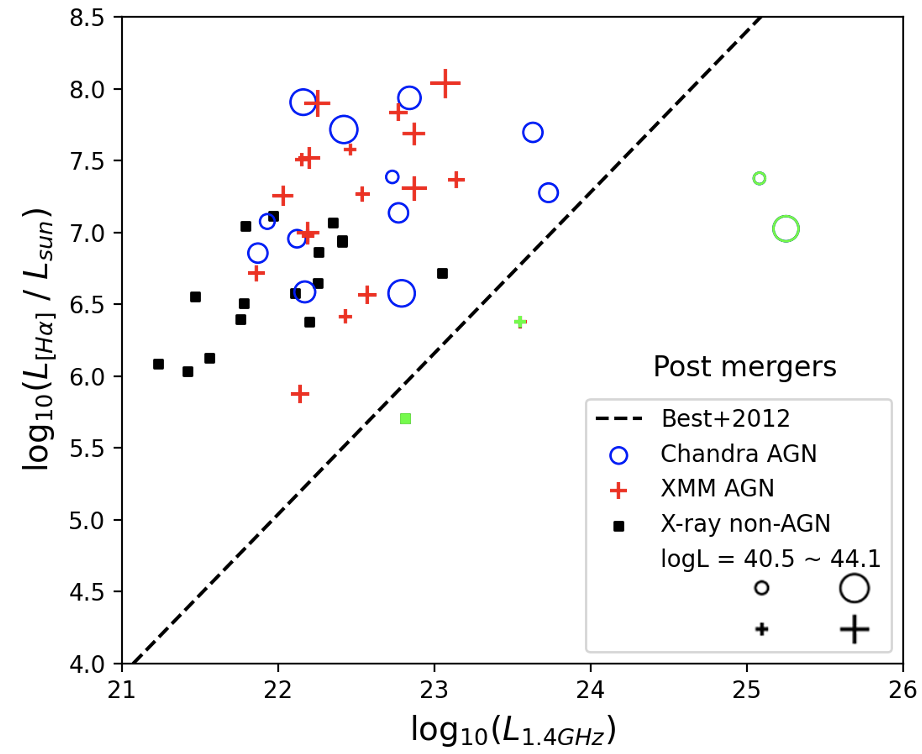}
\end{minipage}
\begin{minipage}{0.45\textwidth}
\includegraphics[width=\linewidth]{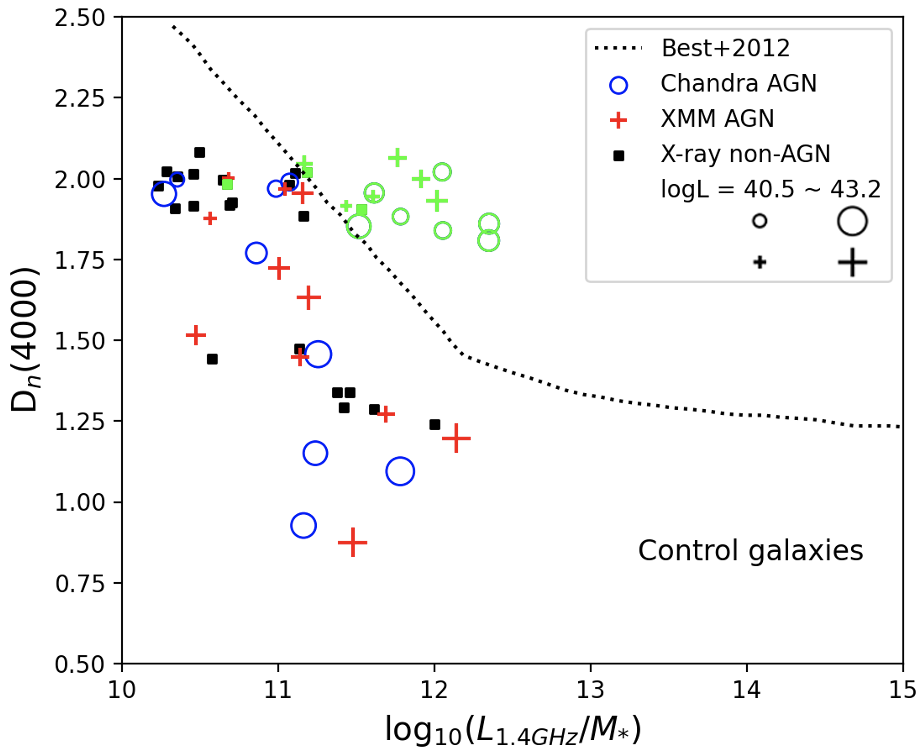}
\end{minipage}
\begin{minipage}{0.45\textwidth}
\includegraphics[width=\linewidth]{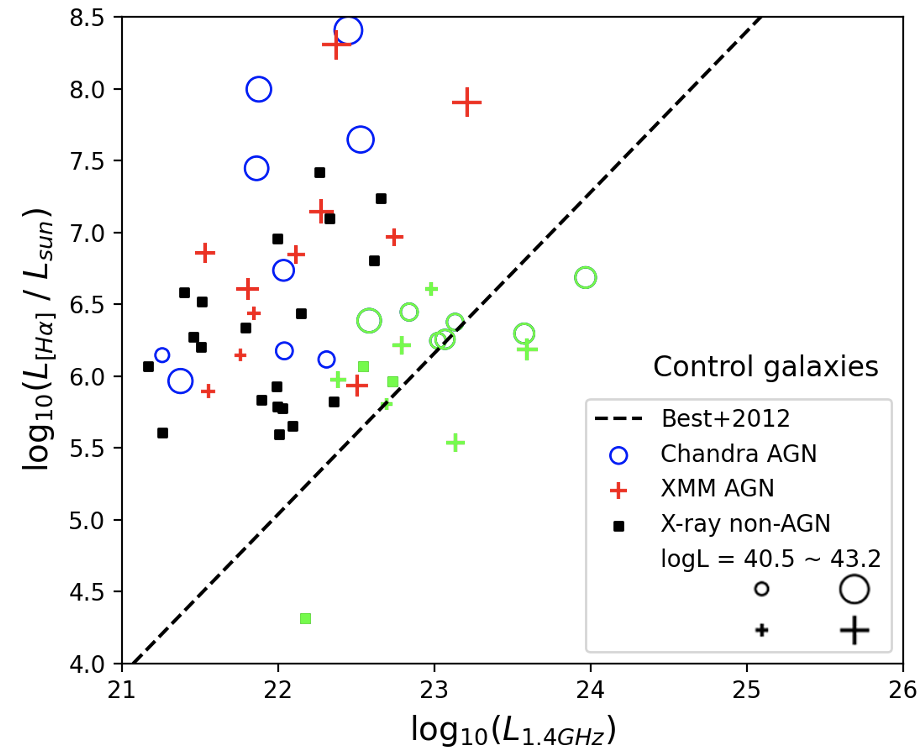}
\end{minipage}
\caption{The $D_n(4000)$ strength vs 1.4 GHz radio luminosity per stellar mass (left panel) and the H$\alpha$ luminosity in solar unit vs 1.4 GHz radio luminosity (right panel) for 77 post-mergers (top) and 233 control galaxies (bottom). The curve on the $D_n(4000)$ vs $L_{1.4GHz}/M_{\ast}$ plane is adopted from \cite{2005MNRAS.362....9B}. For $L_{1.4GHz}/M_{\ast} < 12.2$, a straight-line cut is adopted from \cite{2008MNRAS.384..953K} with the equation $D_n(4000) = 1.45-0.55 \times$ (log $(L_{1.4GHz}/M_{\ast})-12.2)$. The straight line on the $L_{H\alpha}/L_{Sun}$ vs $L_{1.4GHz}$ plane is adopted from \cite{2012MNRAS.421.1569B} with the equation log $(L_{H\alpha}/L_{Sun}) = 1.12 \times$ (log $L_{1.4GHz}-17.5)$. Galaxies with X-ray AGN identified in Chandra are shown as blue circles, while AGN identified in XMM are shown as red crosses. Galaxies not hosting X-ray AGN are shown as squares. In this work, galaxies classified by either one of these two planes are considered to be radio AGN (shown in green). There are four radio AGN classified in post-mergers. Three out of these four are also identified in X-rays. There are 16 radio AGN identified in control galaxies, and 13 out of 16 are also X-ray AGN. }
\label{fig:RadioAGNdiagnostic}
\end{center}
\end{figure*}

\subsection{Radio AGN Diagnostics and Frequency}
There are 77 post-mergers out of the 79 that have been observed by the VLA FIRST survey. Out of the 77, 49 have radio detections showing the presence of a radio core in the center of the galaxies. Radio lobes have also been found in some sources, indicating AGN. However, both AGN and star formation can produce core radio emission. 
In order to exclude galaxies with radio emission dominated by star formation, we apply two methods to identify radio AGN. The first method is to compare the strength of the $D_n(4000)$ break in the optical spectra to the 1.4 GHz radio luminosity per stellar mass (see Figure~\ref{fig:RadioAGNdiagnostic} left). The demarcation curve between radio-loud AGN and normal star-forming galaxies was developed by \cite{2005MNRAS.362....9B} and modified by \cite{2008MNRAS.384..953K}. It is 0.225 higher in $D_n(4000)$ than the track produced by a galaxy with an exponentially declining star formation rate of 3 Gyr e-folding time. Radio AGN are expected to have radio luminosities higher than normal star-forming galaxies (to the right of the curve). 
The second method is to compare the H$\alpha$ luminosity in solar unit to the 1.4 GHz radio luminosity (see Figure~\ref{fig:RadioAGNdiagnostic} right). Galaxies to the right of the straight line show a radio luminosity excess relative to star-forming galaxies, and hence they are identified as radio-loud AGN. 
Figure~\ref{fig:RadioAGNdiagnostic} shows these two planes for the 77 post-merger galaxies in the upper panel and 233 control galaxies in the lower panel. The 1.4 GHz radio luminosities are obtained from the VLA FIRST catalog, while the strength of $D_n(4000)$ and H$\alpha$ luminosities are obtained or derived from the MPA-JHU catalog. Galaxies located either above the division line on the `$D_n(4000)$ vs $L_{1.4GHz}/M_{\ast}$' plane or below the division line on the $L_{H\alpha}/L_{Sun}$ vs $L_{1.4GHz}$ plane are classified as radio AGN, which are shown in green in Figure~\ref{fig:RadioAGNdiagnostic}. Out of the 77 post-mergers, 4 are classified as radio-loud AGN, with a radio AGN fraction of $5.2\% \pm 2.5\%$. 

In the control sample, all the 233 control galaxies are observed by the VLA FIRST survey, out of which, 56 have radio detections showing a radio core. By using the same identification method, 16 radio AGN are identified from the 233 X-ray observed control sample, which yields a radio AGN fraction of $6.9\% \pm 1.7\%$. The AGN fraction in post-mergers and in control galaxies are comparable to each other. Thus, there is no enhancement in the radio AGN fraction in post-mergers relative to control galaxies. This suggests that mergers may have no role in triggering radio AGN. It is interesting to note that the radio AGN fraction obtained is similar to that in higher-redshift studies such as \cite{2009ApJ...696..891H}. Unlike that study, we find most of our radio AGN in both post-merger and control galaxies are identified as X-ray AGN. We discuss this further in Section~4.

\begin{figure*}[t]
\begin{center}
\begin{minipage}{0.45\textwidth}
\includegraphics[width=\linewidth]{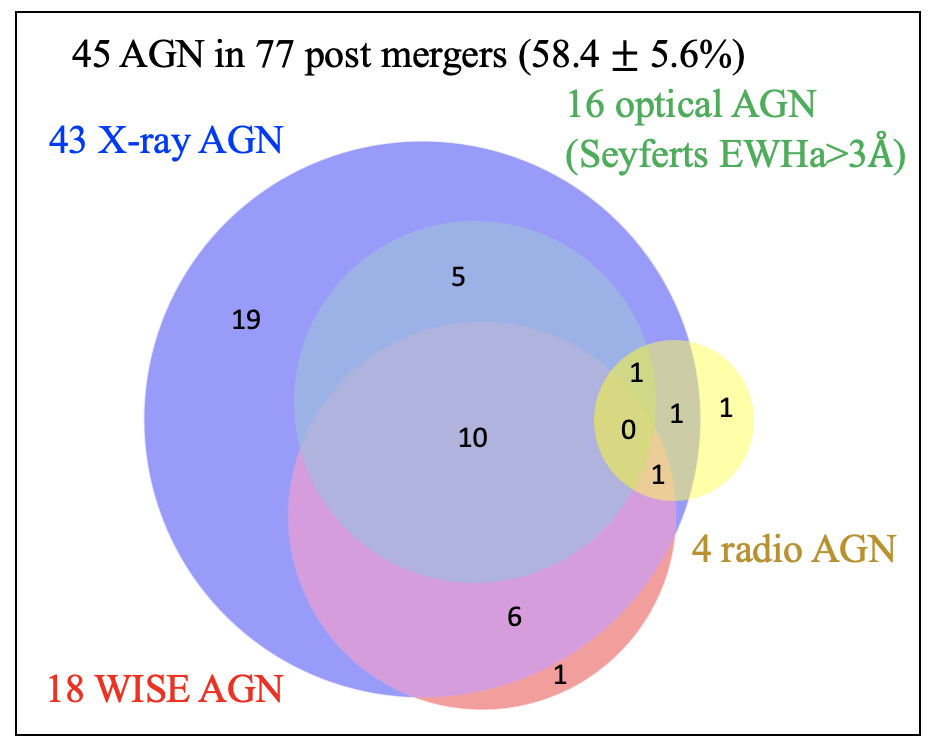}
\end{minipage}
\begin{minipage}{0.45\textwidth}
\includegraphics[width=\linewidth]{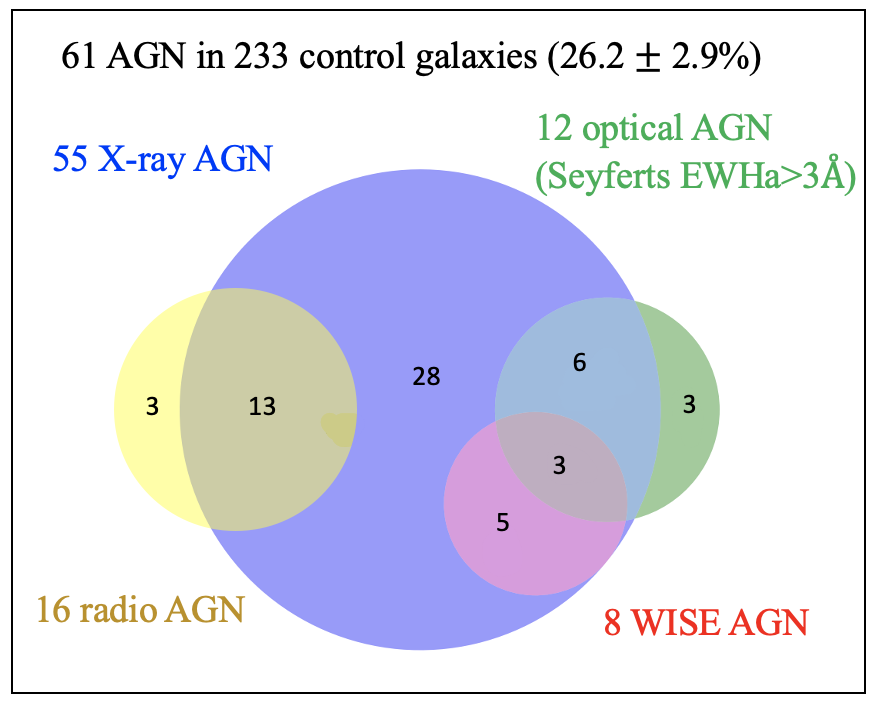}
\end{minipage}
\begin{minipage}{0.45\textwidth}
\includegraphics[width=\linewidth]{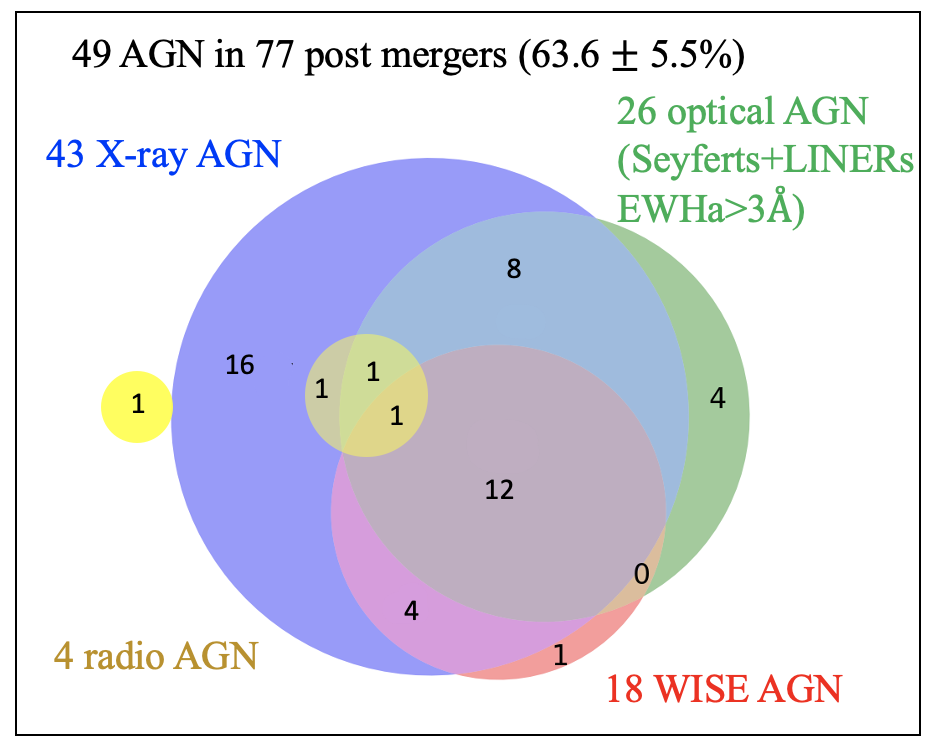}
\end{minipage}
\begin{minipage}{0.45\textwidth}
\includegraphics[width=\linewidth]{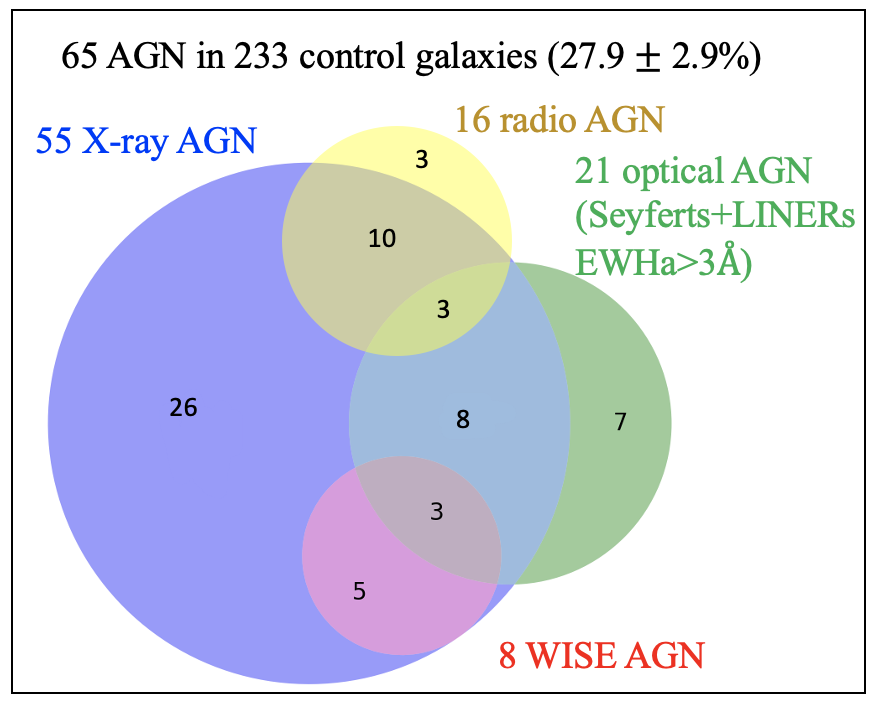}
\end{minipage}
\caption{Venn diagrams for multiwavelength AGN. The upper panel shows the AGN identified in X-rays, infrared, radio, and optical (Seyferts only) for (left) the 77 post-mergers and (right) the 233 control galaxies. The lower panel shows the same four wavelength bands but considers both Seyferts and LINERs as optical AGN. X-ray AGN are shaded in blue. Radio AGN are shaded in yellow, while infrared AGN are in red and optical AGN are in green. The multiwavelength AGN fraction are shown above the diagram. The AGN excess in post-mergers relative to control galaxies is $\sim 2.2$ when considering Seyferts only as optical AGN, and $\sim 2.3$ including LINERs as optical AGN.}
\label{fig:AGNvenndiagram}
\end{center} 
\end{figure*}

\begin{deluxetable*}{|c|c|c|c|}[t]
\tablenum{3}
\tabletypesize{\small}
\tablecaption{Dependence of AGN fraction and excess on AGN diagnostics}
\tablewidth{0pt}
\tablehead{
\colhead{\textbf{ }} & \colhead{AGN in post-mergers} & \colhead{AGN in control galaxies} & \colhead{AGN Excess}
}
\startdata
Chandra & 20 / 38 ($52.6\% \pm 8.1\%$) & 29 / 129 ($22.5\% \pm 3.7\%$) & $2.3 \pm 0.5$ \\
XMM & 24 / 41 ($58.5\% \pm 7.7\%$) & 26 / 104 ($25.0\% \pm 4.2\%$) & $2.3 \pm 0.5$ \\
Chandra+XMM & 44 / 79 ($55.7\% \pm 5.6\%$) & 55 / 233 ($23.6\% \pm 2.8\%$) & $2.4 \pm 0.4$ \\
WISE infrared & 18 / 79 ($22.8\% \pm 4.7\%$) & ~~8 / 233 ($~3.4\% \pm 1.2\%$) & $6.7 \pm 2.7$ \\
FIRST radio & ~4 / 77 ($~~5.2\% \pm 2.5\%$) & 16 / 233 ($~6.9\% \pm 1.7 \%$) & $0.8 \pm 0.4$ \\
Optical (BPT, Seyfert) & 16 / 79 ($20.3\% \pm 4.5\%$) & 14 / 233 ($~6.0\% \pm 1.6\%$) & $3.4 \pm 1.2$ \\
Optical (BPT, Seyfert, EWH$\alpha>3$\AA) & 16 / 79 ($20.3\% \pm 4.5\%$) & 12 / 233 ($~5.2\% \pm 1.4\%$) & $3.9 \pm 1.4$ \\
Optical (BPT, Seyfert+LINER) & 40 / 79 ($50.6\% \pm 5.6\%$) & 64 / 233 ($27.5\% \pm 2.9\%$) & $1.8 \pm 0.3$ \\
Optical (BPT, Seyfert+LINER, EWH$\alpha>3$\AA) & 27 / 79 ($34.2\% \pm 5.3\%$) & 21 / 233 ($~9.0\% \pm 1.9\%$) & $3.8 \pm 1.0$ \\
Optical (BPT, S+L+composite) & 66 / 79 ($83.5\% \pm 4.2\%$) & 125 / 233 ($53.6\% \pm 3.3\%$) & $1.6 \pm 0.1$ \\
Optical (BPT, S+L+composite, EWH$\alpha>3$\AA) & 52 / 79 ($65.8\% \pm 5.3\%$) & 55 / 233 ($23.6\% \pm 2.8\%$) & $2.8 \pm 0.4$ \\
Optical (WHaN, strong AGN) & 45 / 79 ($57.0\% \pm 5.6\%$) & 34 / 233 ($14.6\% \pm 2.3\%$) & $ 3.9 \pm 0.7$ \\
Optical (WHaN, strong+weak AGN) & 51 / 79 ($64.6\% \pm 5.4\%$) & 53 / 233 ($22.7\% \pm 2.7\%$) & $2.8 \pm 0.4$ \\
X-ray, WISE, radio & 45 / 77 ($58.4\% \pm 5.6\%$) & 58 / 233 ($24.9\% \pm 2.8\%$) & $2.3 \pm 0.3$ \\
WISE, radio, optical (S+L, EWH$\alpha>3$\AA) & 33 / 77 ($42.9\% \pm 5.6\%$) & 39 / 233 ($16.7\% \pm 2.4\%$) & $2.6 \pm 0.5$ \\
X-ray, WISE, radio, optical (Seyferts,EWH$\alpha>3$\AA) & 45 / 77 ($58.4\% \pm 5.6\%$) & 61 / 233 ($26.2\% \pm 2.9\%$) & $2.2 \pm 0.3$ \\
X-ray, WISE, radio, optical (S+L,EWH$\alpha>3$\AA) & 49 / 77 ($63.6\% \pm 5.5\%$) & 65 / 233 ($27.9\% \pm 2.9\%$) & $2.3 \pm 0.3$ \\
X-ray, WISE, radio, optical (S+L+composite,EWH$\alpha>3$\AA) & 59 / 77 ($76.6\% \pm 4.8\%$) & 91 / 233 ($39.1\% \pm 3.2\%$) & $2.0 \pm 0.2$ \\
X-ray, WISE, radio, optical (WhaN) & 60 / 77 ($77.9\% \pm 4.7\%$) & 99 / 233 ($42.5\% \pm 3.2\%$) & $1.8 \pm 0.2$ 
\enddata
\tablecomments{The table lists the number of AGN and the AGN fraction in post-mergers and control galaxies and the AGN excess in different wavelength bands. The AGN excess is defined as the AGN fraction in post-mergers relative to that in control galaxies. When calculating the AGN fraction for a specific wavelength band or telescope, only the galaxies with existing observations are taken into account. The WHaN AGN fraction \citep{2011MNRAS.413.1687C}, described in the Appendix, is also shown in this table.}
\end{deluxetable*}

\section{Discussion}
\subsection{Multiwavelength AGN Fraction}
In the previous section, we investigated the merger–AGN relation in post-merger galaxies and controls individually in the X-rays, optical, mid-infrared, and radio. We found a significant AGN excess in post-mergers compared to control galaxies in the X-ray, infrared, and optical bands. In this section, we will discuss the multiwave-band AGN fraction by combing all four AGN identification methods. We identify a galaxy as an AGN if it is detected as an AGN in at least one diagnostic. We will first calculate the multiwave-band AGN fraction by considering only Seyferts as optical AGN. Then we will include LINERs. We will also apply the equivalent width cut on H$\alpha$ to be $>$ 3\AA.

Figure~\ref{fig:AGNvenndiagram} shows Venn diagrams of AGN identified in all four wavebands: X-rays (blue), mid-IR (red), radio (yellow), and optical (green) for post-mergers (left) and their control galaxies (right). The top panels consider Seyferts only as optical AGN while the bottom panel considers both Seyferts and LINERs as optical AGN. 
Note that only 77 of the 79 post-mergers are shown in these diagram as two post-mergers do not have FIRST radio observations.
The upper panels in Figure~\ref{fig:AGNvenndiagram} show that 45 of the 77 post-mergers ($58.4\% \pm 5.6 \%$) are identified as AGN in at least one band (X-rays, IR, radio and optical) while 61 of the 233 control galaxies ($26.2\% \pm 2.9\%$) are identified as AGN. 
Nearly all WISE mid-IR and FIRST radio AGN are also X-ray AGN no matter if they are post-mergers or control galaxies. This is also true for a large fraction of optically identified AGN.
Thus the overall multiwavelength AGN fraction is dominated by the X-ray-detected sample.
The multiwavelength AGN fraction in post-merger galaxies is $\sim 2.2$ times higher than in noninteracting control galaxies.
By applying a hypergeometric test, we can rule out that post-mergers and control galaxies have similar AGN fractions at $>6.5\sigma$ confidence level. Applying a Fisher exact test, we find post-mergers and AGN are strongly correlated with a p-value of $10^{-7}$.

The bottom Venn diagrams in Figure~\ref{fig:AGNvenndiagram} show the multiwavelength AGN (including LINERs as optical AGN) identified in post-mergers and control galaxies. There are 49 AGN out of 77 post-mergers identified in at least one band. The multiwavelength AGN fraction including optical LINERs in the post-merger sample is $63.6\% \pm 5.5\%$. In the control sample, 65 AGN out of 233 control galaxies are identified in at least one band, which is shown as the last Venn diagram in Figure~\ref{fig:AGNvenndiagram}. This implies an AGN fraction of $27.9\% \pm 2.9\%$ in control galaxies. The AGN excess in post-mergers is $\sim 2.3$ relative to control galaxies.  By applying a similar hypergeometric test, we can rule out that post-mergers and control galaxies have similar AGN fractions at $>6.5\sigma$ confidence level. A Fisher exact test shows that mergers and AGN are strongly related with a p-value = $10^{-8}$ for the null hypothesis. 

We find that a large fraction of AGN are identified mainly in X-rays ($\sim80\%$ in both post-mergers and control galaxies). In fact, $\sim$ 40\% of AGN are only identified in deep X-ray imaging for both post-mergers and control galaxies. If we consider AGN classified in WISE, radio, and optical, the multiwavelength AGN fraction is $42.9\% \pm 5.6\%$ in post-mergers and $16.7\% \pm 2.4\%$ in control galaxies. Although it still implies a similar AGN excess (2.6), the multiwavelength AGN fraction is lower without X-ray AGN. Hence, deep X-ray observations are crucial to probe the complete demographics of AGN and the merger–AGN relation.

Table 3 summarizes the AGN fractions and excesses obtained using different AGN selection techniques for both post-mergers and their controls. In every wavelength range, except radio, post-mergers have a higher AGN fraction compared to control galaxies. However, the exact values of the AGN fraction and excess obtained differ based on the diagnostic. 

\begin{figure}
\begin{center}
\begin{minipage}{0.47\textwidth}
\includegraphics[width=\linewidth]{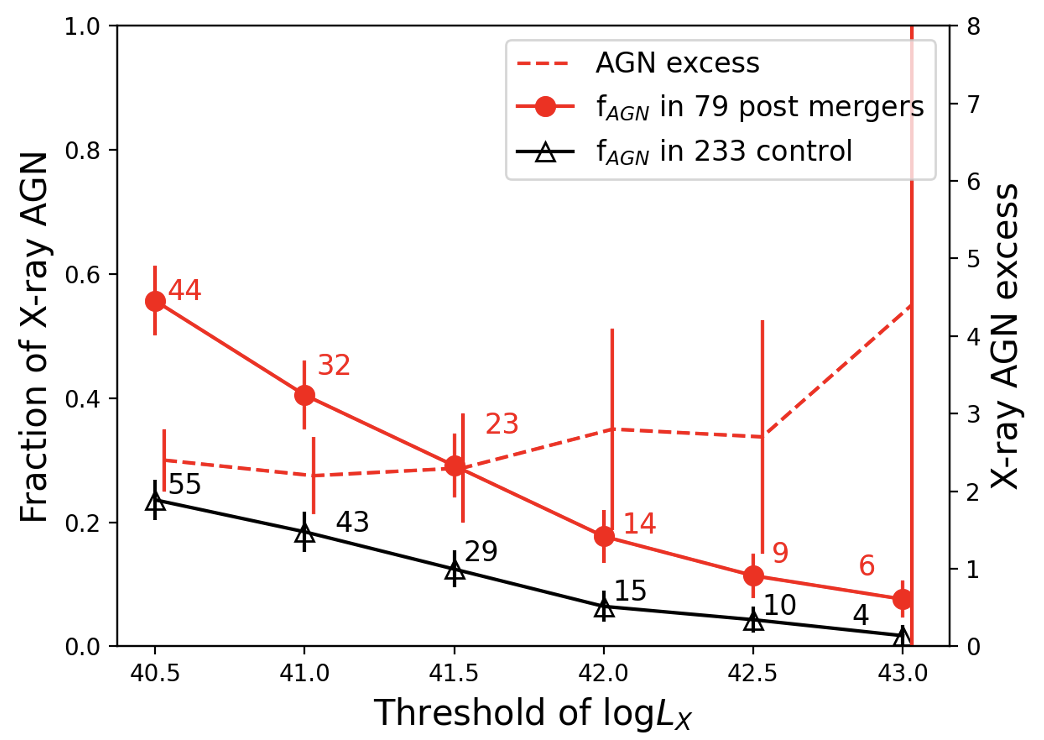}
\end{minipage}
\caption{X-ray AGN fraction in 79 post-mergers (red) and 233 control galaxies (black) as a function of the X-ray luminosity threshold. The numbers of X-ray AGN above each luminosity threshold are shown next to the data points. The red dashed line shows the X-ray AGN excess as a function of the X-ray luminosity threshold.}
\label{fig:f_AGN_Lx}
\end{center} 
\end{figure}

\begin{figure*}[t!]
\begin{center}
\begin{minipage}{0.32\textwidth}
\includegraphics[width=\linewidth]{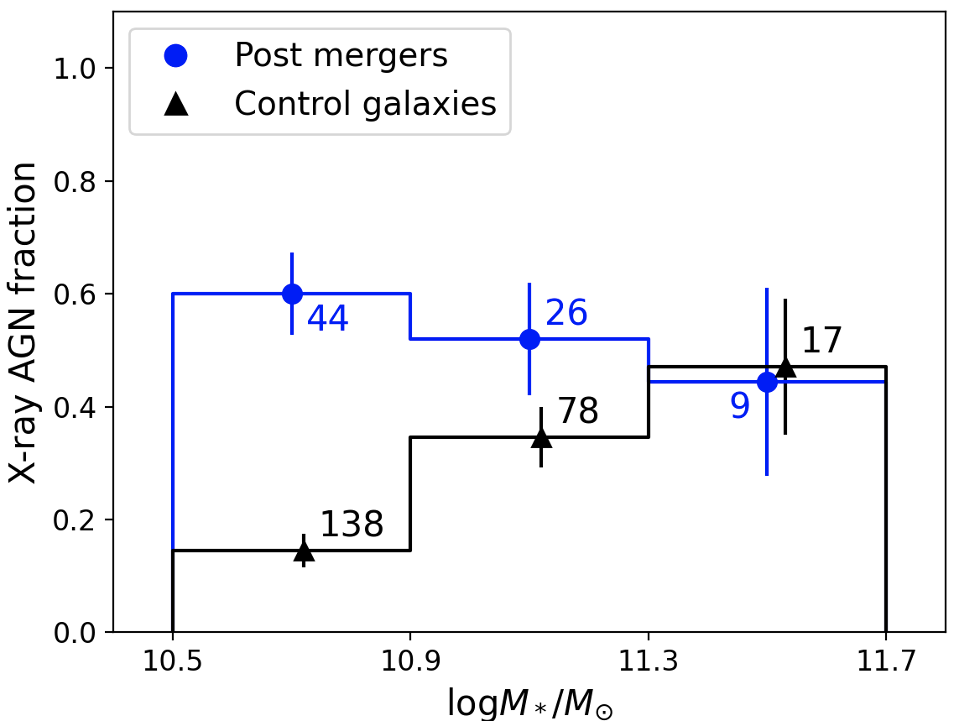}
\end{minipage}
\begin{minipage}{0.32\textwidth}
\includegraphics[width=\linewidth]{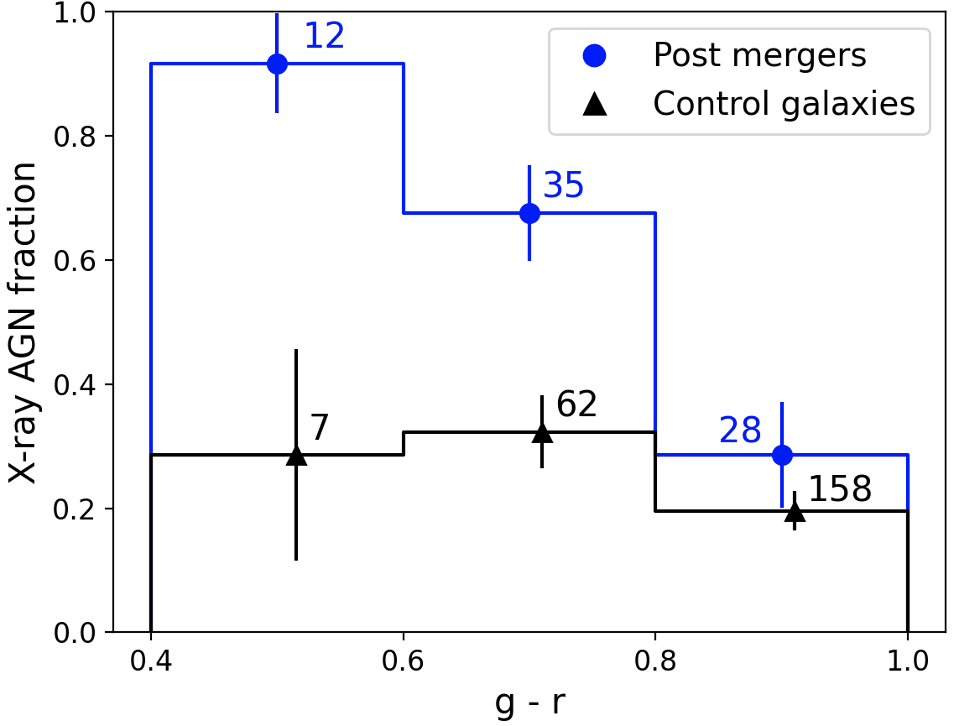}
\end{minipage}
\begin{minipage}{0.32\textwidth}
\includegraphics[width=\linewidth]{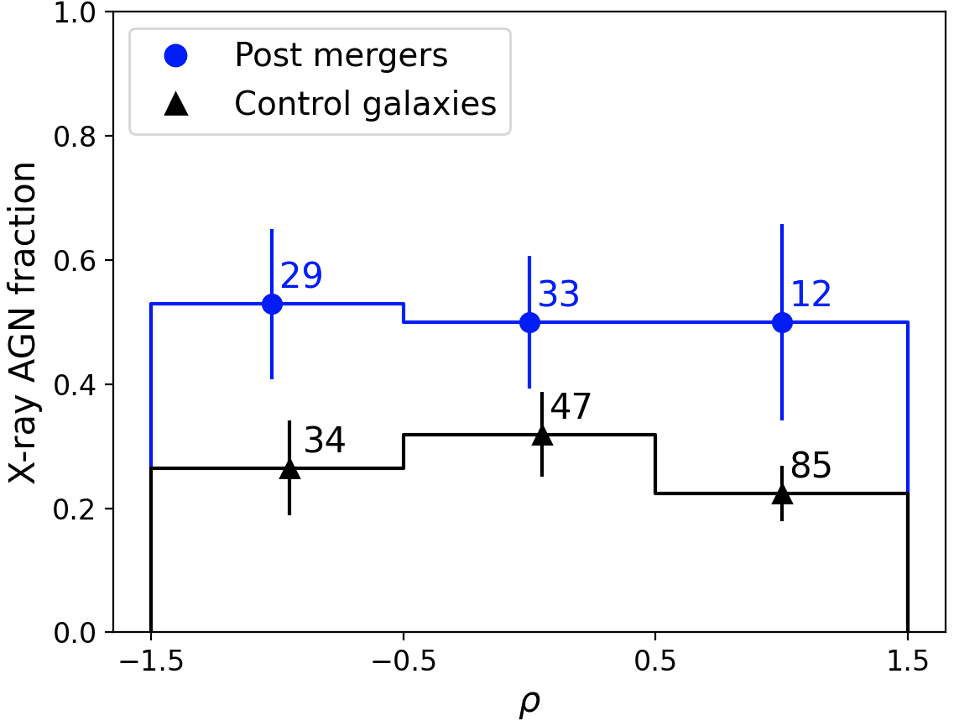}
\end{minipage}
\begin{minipage}{0.32\textwidth}
\includegraphics[width=\linewidth]{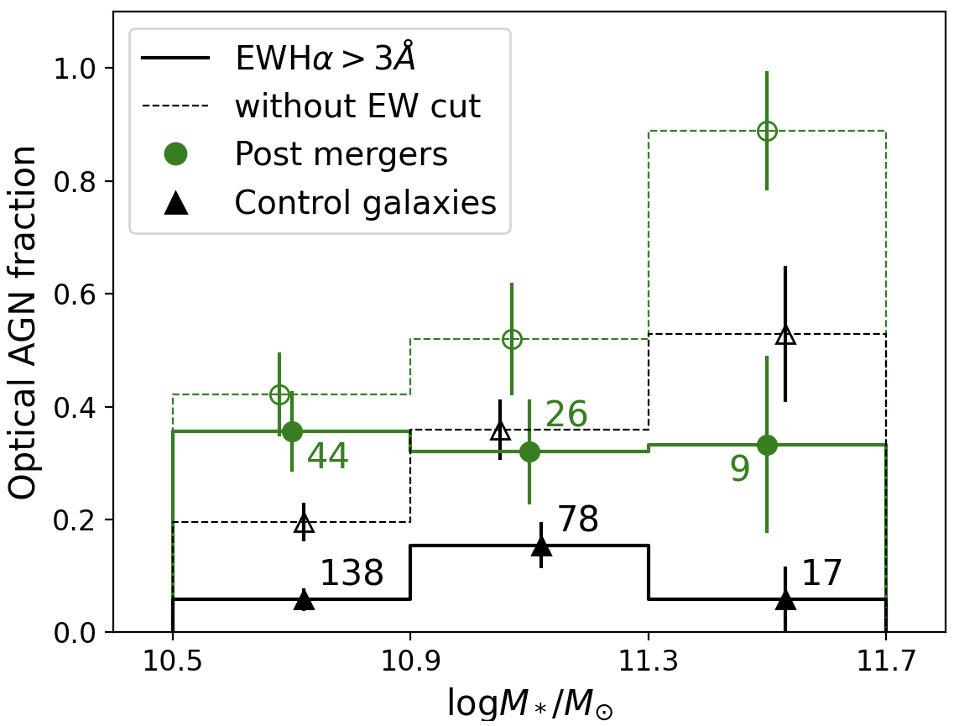}
\end{minipage}
\begin{minipage}{0.32\textwidth}
\includegraphics[width=\linewidth]{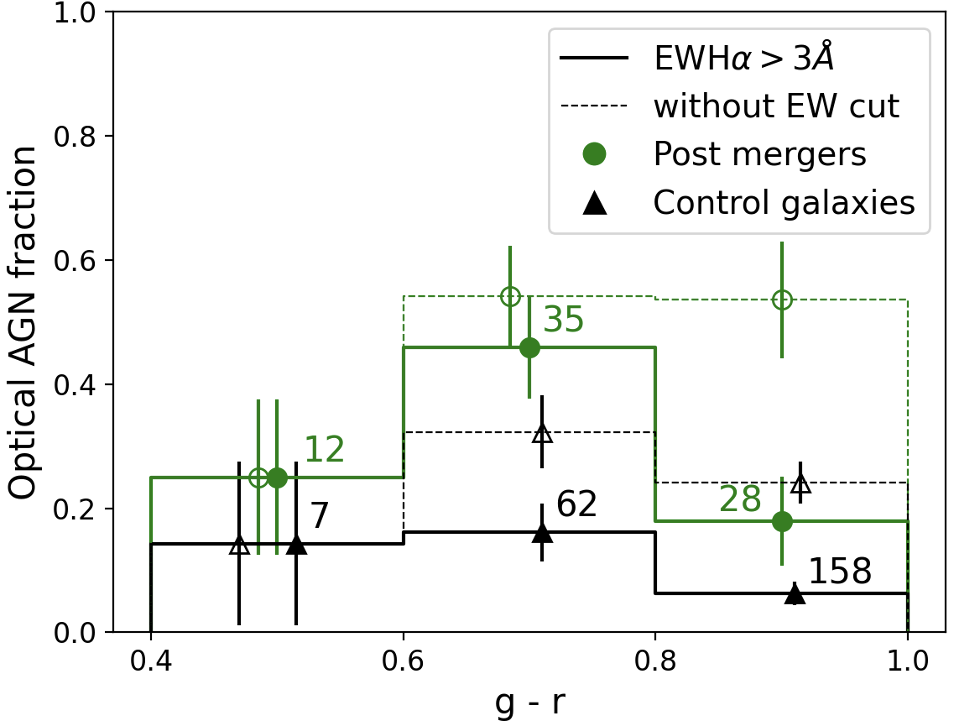}
\end{minipage}
\begin{minipage}{0.32\textwidth}
\includegraphics[width=\linewidth]{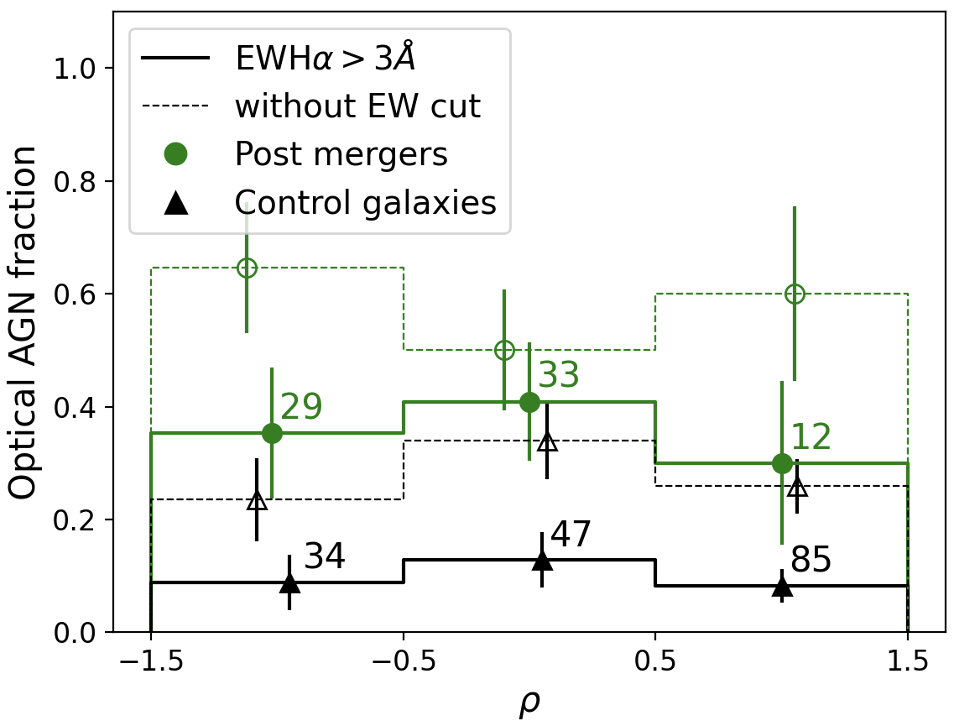}
\end{minipage}
\begin{minipage}{0.32\textwidth}
\includegraphics[width=\linewidth]{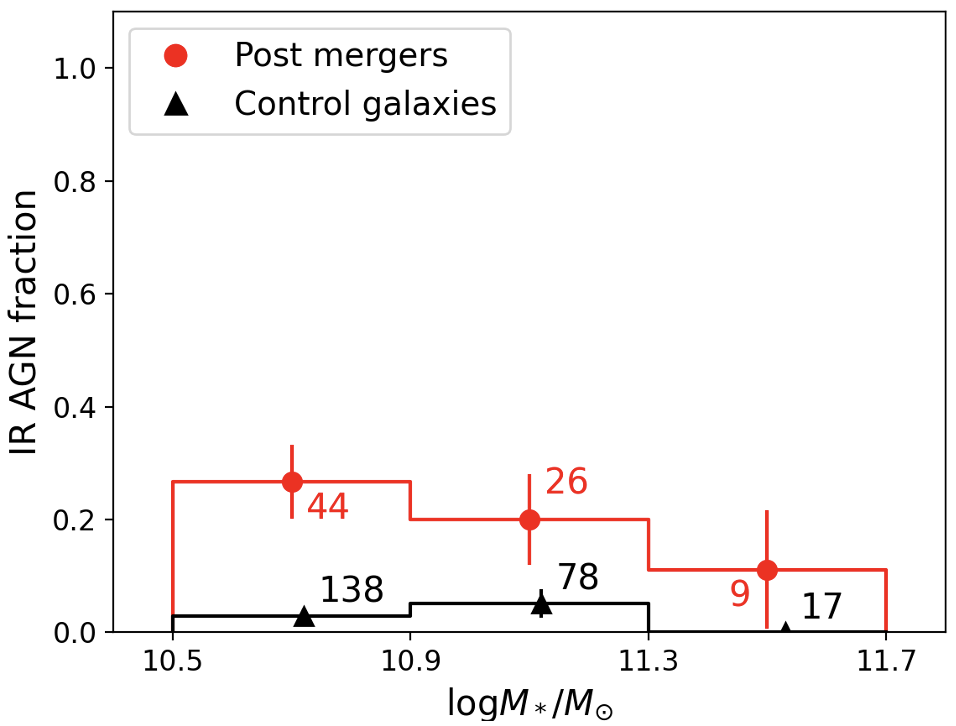}
\end{minipage}
\begin{minipage}{0.32\textwidth}
\includegraphics[width=\linewidth]{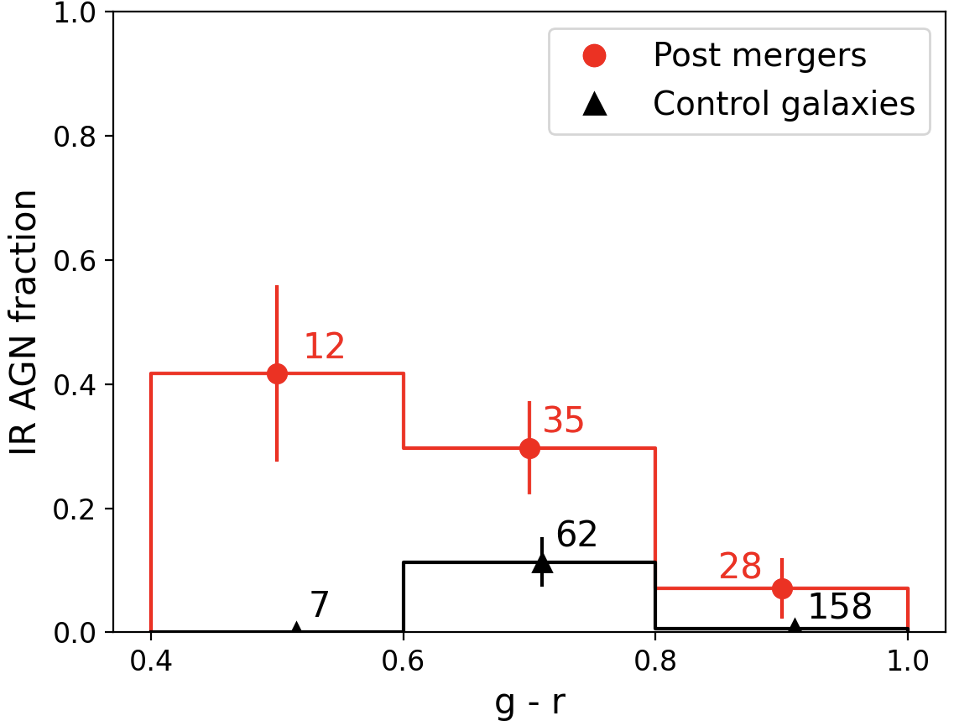}
\end{minipage}
\begin{minipage}{0.32\textwidth}
\includegraphics[width=\linewidth]{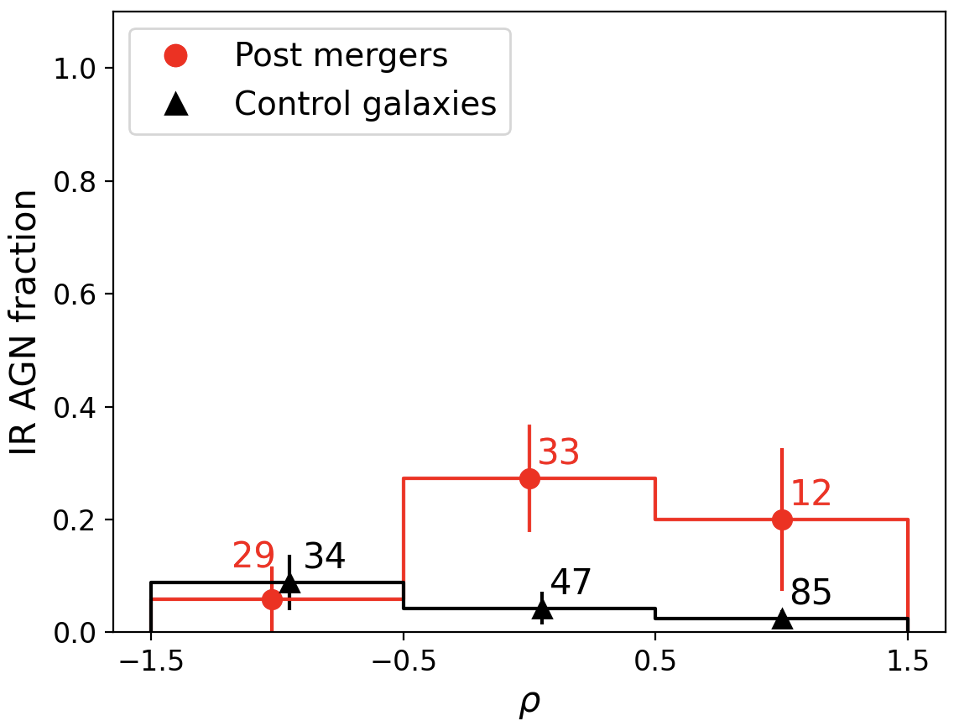}
\end{minipage}
\caption{X-ray AGN fraction (blue circles), optical AGN fraction (green circles), and IR AGN fraction (red circles) in post-mergers and control galaxies (black triangles) as a function of the stellar mass, color, and environment density. Here we only consider X-ray AGN with luminosities $>=10^{40.5}$ erg s$^{-1}$. In the optical, we consider Seyferts + LINERs with EW(H$\alpha$)$>$3\AA\ as AGN. We also show the Seyferts + LINERs without the equivalent width cut with dashed lines. The numbers of galaxies per bin are shown near the points. There is an increasing trend of the X-ray AGN fraction in control galaxies as a function of the stellar mass while the X-ray AGN fraction in post-mergers stays roughly the same. The optical and IR AGN fractions do not show a dependence on the stellar mass in neither post-mergers nor controls. There is a decreasing trend of the X-ray and IR AGN fractions as a function of the color in post-mergers while the AGN fraction in control galaxies does not show a dependence on color. The AGN fraction stays roughly consistent in different environment densities in both post-mergers and control galaxies in all three wavebands.}
\label{fig:agnproperties}
\end{center} 
\vspace{0.5cm}
\end{figure*}

Our multiwavelength AGN excess in optical and WISE diagnostics is consistent with that of many previous works \citep{2011MNRAS.418.2043E, 2013MNRAS.435.3627E, 2014MNRAS.441.1297S, 2017MNRAS.464.3882W, 2018PASJ...70S..37G, 2020A&A...637A..94G}. However, our X-ray analysis is not consistent with a similar work. \cite{2020MNRAS.499.2380S} investigated the AGN fraction in 43 post-mergers by using XMM-Newton observations. They found that $4.7^{+9.3}_{-3.8}\%$ (2/43) of their post-mergers host an X-ray AGN while $2.1^{+1.5}_{-1.0}\%$ (9/430) of noninteracting galaxies have X-ray AGN detections, with an excess of $\sim 2.22^{+4.44}_{-2.22}$ which though comparable to our measured excess was not statistically significant. Compared to their work, we find both a higher X-ray AGN fraction in post-mergers ($55.7\%\pm5.6\%$) and control galaxies ($23.6\%\pm2.8\%$) and a significant AGN excess in post-mergers.  As can be seen in Figure~\ref{fig:LehmerRelation}, most of the X-ray AGN in both our post-merger sample and control sample have luminosities $<10^{43}$ erg s$^{-1}$, with a median value of log L = $41.6$ in post-mergers and log L = $41.4$ in control galaxies. The \cite{2020MNRAS.499.2380S} sample, while not having a single detection threshold, was restricted to objects with luminosities between $10^{41} - 10^{43}$ erg s$^{-1}$, with a median luminosity of $\sim 10^{41.5}$ erg s$^{-1}$. If we restrict our sample to the same luminosity range, we obtain an X-ray AGN fraction of $29.1\% \pm 5.1\%$ in post-mergers and $13.5\% \pm 1.7\%$ in control galaxies leading to an AGN excess of $2.2 \pm 0.5$. We attribute the lower X-ray AGN fraction in \cite{2020MNRAS.499.2380S} to (a) their sample not having uniform depth, (b) a smaller sample size spanning a wider range in merger stages, and (c) having contamination from noninteracting galaxies (like outer ring galaxies) in their post-merger sample.

Figure~\ref{fig:f_AGN_Lx} shows the X-ray AGN fraction (solid line) and AGN excess (dashed line) versus the threshold X-ray luminosity for post-mergers (red lines) and control galaxies (black line). We find that the AGN fraction obtained decreases with increasing X-ray-detection threshold, as expected. However, the AGN excess stays in the range of 2 - 3 for any X-ray threshold luminosity chosen. There are indications that the AGN excess in post-mergers may increase at higher luminosities. However, we do not have the statistics to confirm this. This is consistent with some previous studies that found that major mergers tend to trigger the most luminous AGN more frequently than normal galaxies \citep{2012ApJ...758L..39T, 2015ApJ...806..218G, 2015ApJ...804...34H, 2018MNRAS.480.3562D, 2018PASJ...70S..37G, 2018MNRAS.476.2308W, 2020MNRAS.494.5713M}. However, there are other studies finding no trend between mergers and luminous AGN \citep{2014MNRAS.439.3342V, 2019ApJ...882..141M}. 

Overall, unlike many previous studies mainly focusing on luminous AGN (e.g., \citealp{ 2011ApJ...743....2S, 2014AJ....148..137L, 2017MNRAS.464.3882W, 2018PASJ...70S..37G}), our sample probes AGN with a wider range of X-ray luminosities from $10^{44}$ erg s$^{-1}$ down to $10^{40.5}$ erg s$^{-1}$. By specifically including LLAGN, we are able to have a more complete census of the AGN population. 

\subsection{Dependence of the AGN Fraction on Galaxy Properties}
In addition to AGN selection techniques, sample selection could play a role in understanding the merger–AGN connection. In Figure~\ref{fig:agnproperties}, we investigate the X-ray AGN fraction (top), optical AGN fraction (middle), and WISE IR AGN fraction (bottom) as a function of the stellar mass (left), g - r color (middle), and environment density (right) of the post-mergers and control galaxies. Again, we only consider X-ray AGN with luminosities $>=10^{40.5}$ erg s$^{-1}$. We find that the X-ray AGN fraction in the 79 post-mergers is $\sim 56\%$ and nearly independent of the stellar mass given our measurement errors (blue, solid line), while in the 233 galaxy control sample (black, solid line), the X-ray AGN fraction increases with the stellar mass. This suggests that X-ray LLAGN in post-merger galaxies are a better tracer of the supermassive black hole (SMBH) occupation fraction compared to normal, noninteracting galaxies.

In Figure~\ref{fig:agnproperties} (left, middle row), we consider Seyferts and LINERs with EW(H$\alpha$) $>$ 3\AA\ as AGN (solid lines). We also show the Seyferts + LINERs without the equivalent width cut (dotted lines) for comparison. The optical AGN fraction does not depend on the stellar mass in neither post-mergers nor the control sample. Removing the equivalent width cut in H$\alpha$ leads to a dependence of the AGN fraction on the stellar mass such that the AGN fractions are higher in more massive galaxies. This is likely due to increased contamination by non-AGN LINERs at higher masses. The IR AGN fraction is also consistent with having no dependence on the stellar mass for post-mergers given the error bars.

The middle column in Figure~\ref{fig:agnproperties} shows the dependence of the AGN fraction on the galaxy color. We find a clear trend of decreasing X-ray AGN fraction from blue to red (g - r) colors for post-mergers. 
Bluer post-mergers tend to have a very high X-ray AGN fraction ($\sim 90\%$) than redder post-mergers ($\sim 30\%$). Our control galaxy sample shows no such trend in the X-rays. This trend with the color could be related to the gas content in these merger systems. 

The optical AGN fraction shows no clear dependence on the galaxy color given the error bars although post-mergers with intermediate colors (g-r = 0.6-0.8) may have a slightly higher optical AGN fraction. 

The IR AGN fraction in post-mergers shows a similar dependence on the color as the X-ray AGN fraction such that bluer post-merger galaxies have a higher IR AGN fraction compared to redder galaxies. This is consistent with \cite{2009ApJ...696..891H} finding that AGN identified in the mid-IR tend to be located in blue star-forming galaxies. The control sample shows a slight IR AGN excess for intermediate colors.

\begin{figure*}[t!]
\begin{center}
\begin{minipage}{0.48\textwidth}
\includegraphics[width=\linewidth]{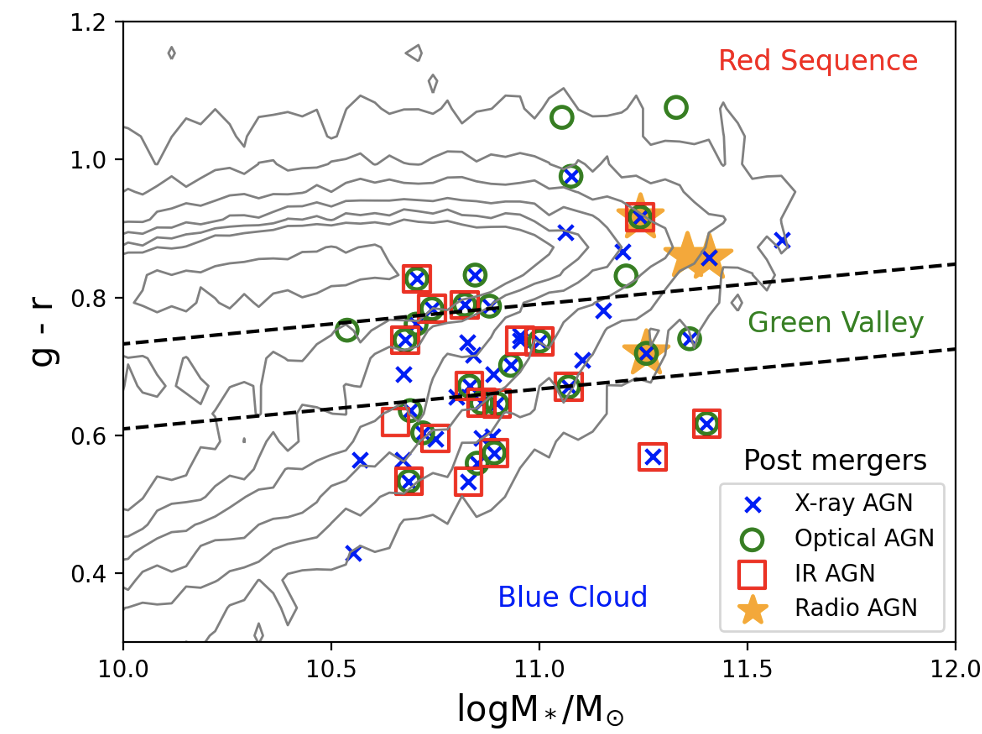}
\end{minipage}
\begin{minipage}{0.48\textwidth}
\includegraphics[width=\linewidth]{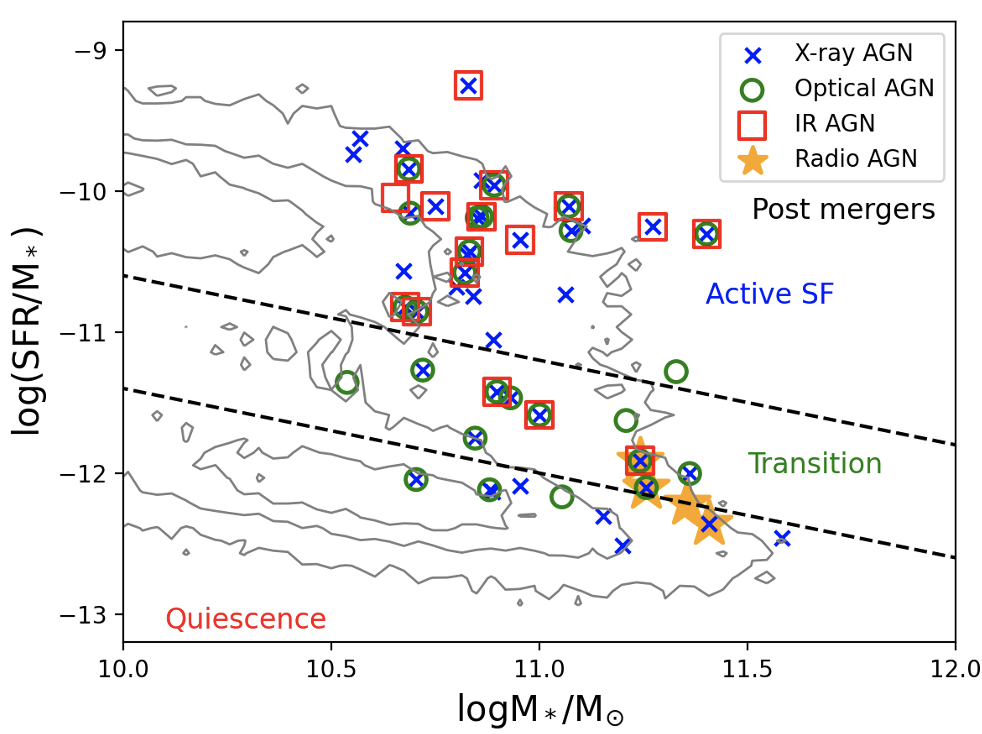}
\end{minipage}
\begin{minipage}{0.48\textwidth}
\includegraphics[width=\linewidth]{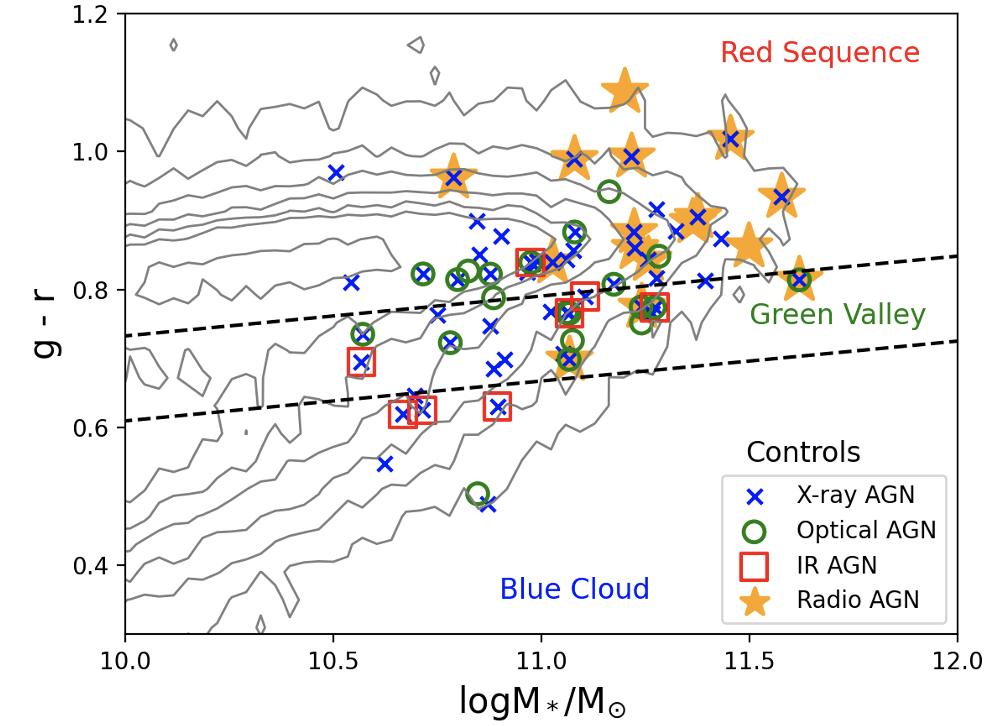}
\end{minipage}
\begin{minipage}{0.48\textwidth}
\includegraphics[width=\linewidth]{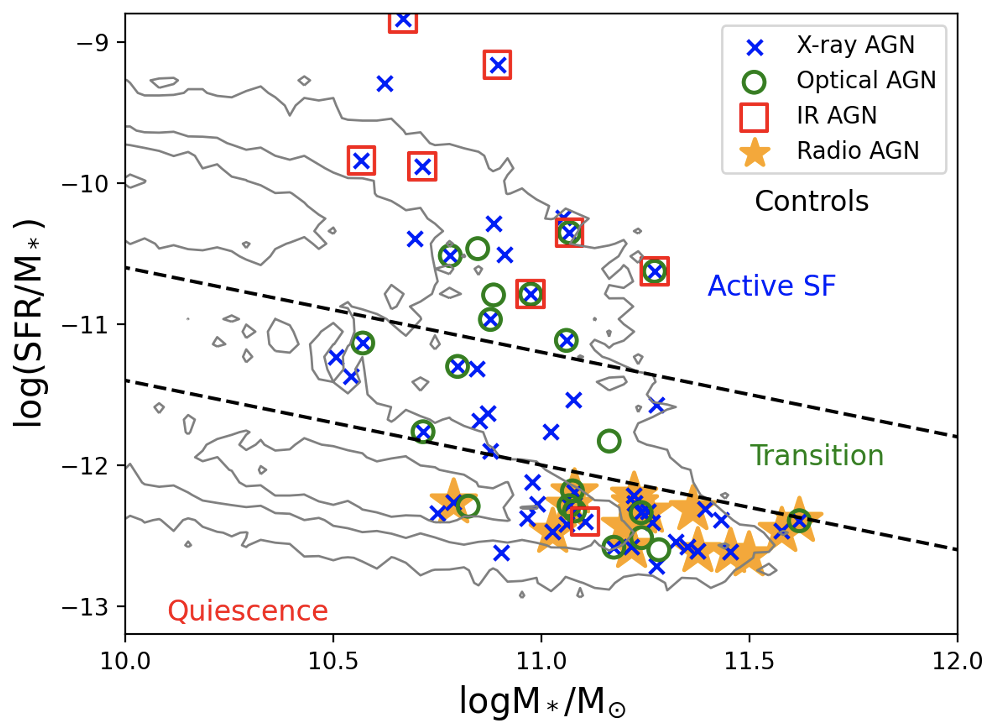}
\end{minipage}
\caption{The (g - r) color vs stellar mass diagram (left) and the specific SFR vs stellar mass diagram (right) of AGN in post-mergers on the upper panel and control galaxies on the lower panel. AGN identified in X-rays are shown as blue crosses; in optical as green, open circles; in WISE mid-IR as red, open boxes; and in radio as filled, orange stars. The grey contours show the N23 SDSS sample of 113K galaxies. The outermost grey contours include 99\% of all galaxies in the N23 sample from SDSS DR14. The (g - r) color vs stellar mass diagram is divided into red-sequence, green-valley, and blue-cloud regions by dashed lines. The specific SFR vs stellar mass diagram is divided into active star-forming, transition, and quiescent regions. The black dashed lines are drawn by eye. Optical and X-ray sources span the entire range of galaxy properties. Radio AGN are in the red-sequence/quiescent region while IR AGN are in the blue-cloud/active SF region and green-valley/transition region.
}
\label{fig:colormass}
\end{center} 
\end{figure*}

When constructing the control sample, we only controlled for the stellar mass and redshift.
However, the environment of galaxies can affect the gas content and triggering of an AGN \citep{2020MNRAS.491.4045K}. Figure~\ref{fig:agnproperties} (right) shows the dependence of the X-ray AGN fraction on the environment density in post-mergers and the control sample. The environment density for our galaxies is obtained from \cite{2006MNRAS.373..469B}. We find that the X-ray AGN fraction in both post-merger and control galaxies is nearly independent of the environment. The optical and IR AGN fractions also show no dependence (or a very mild dependence) on the galaxy environment within errors. It should be noted that our sample does not probe the most massive clusters. Hence, we cannot rule out a dependence of the AGN fraction on the environment in the densest regions.

The color and stellar mass of our sample are key factors in determining the AGN fraction in both post-mergers and controls. Figure~\ref{fig:colormass} shows the (g - r) color vs stellar mass diagram on the left and the specific SFR vs stellar mass diagram on the right for AGN identified in post-mergers (top) and controls (bottom). The gray contours are for the entire N23 SDSS 113K galaxy sample. X-ray-identified AGN are shown as blue crosses, optical AGN are shown as green open circles, IR AGN are shown as red open squares, and radio AGN are shown as yellow stars. The dashed lines in each plot show the demarcation between passive/red-sequence, green-valley, and star-forming/blue-cloud galaxies. We find that X-ray AGN are distributed throughout the color–mass and sSFR–mass space for both post-mergers and controls. The same is true for optical AGN. IR AGN are more likely to be located in star-forming/blue-cloud or green-valley galaxies. On the contrary, AGN identified in the radio tend to appear in the passive/red-sequence regions. This is consistent with \cite{2009ApJ...696..891H} and numerous studies (\citealp{2022A&ARv..30....6M} and references therein) which claim that radio AGN are more likely in massive, red galaxies where they are important in maintaining quenching. The mechanisms triggering radio AGN are expected to be related to the stellar mass of the galaxy \citep{2022A&ARv..30....6M} and not external triggers.

\begin{figure*}[t!]
\begin{center}
\begin{minipage}{0.32\textwidth}
\includegraphics[width=\linewidth]{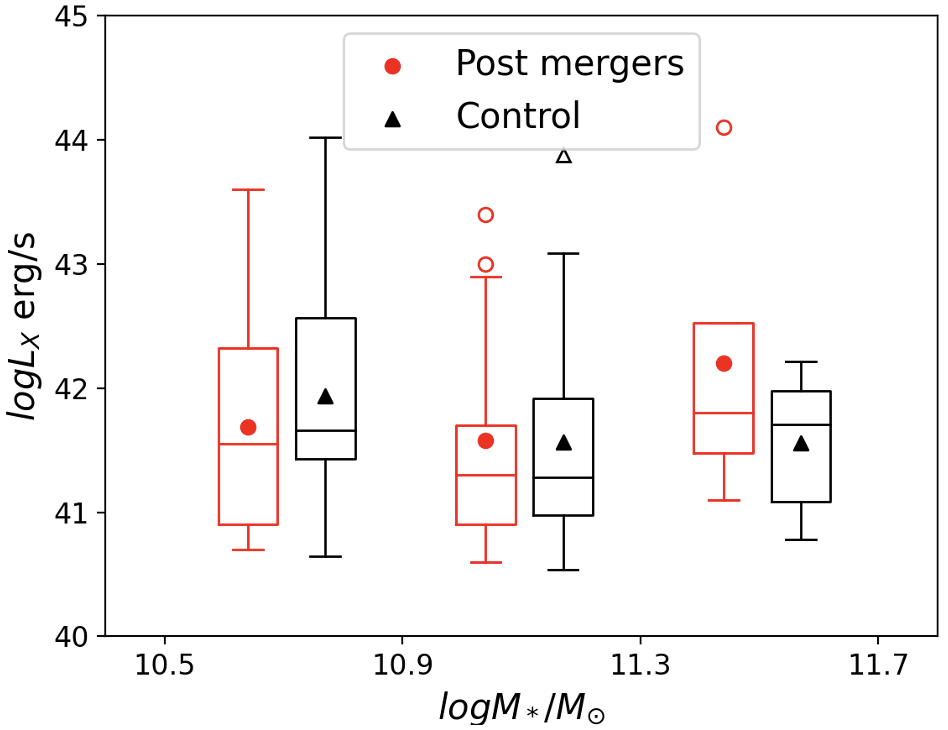}
\end{minipage}
\begin{minipage}{0.32\textwidth}
\includegraphics[width=\linewidth]{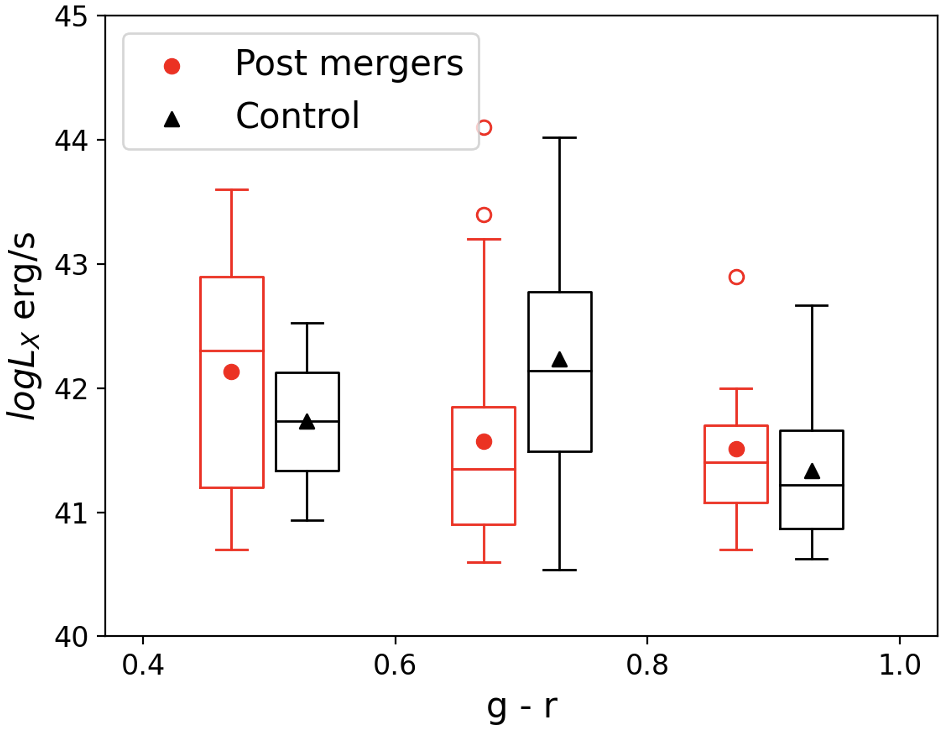}
\end{minipage}
\begin{minipage}{0.32\textwidth}
\includegraphics[width=\linewidth]{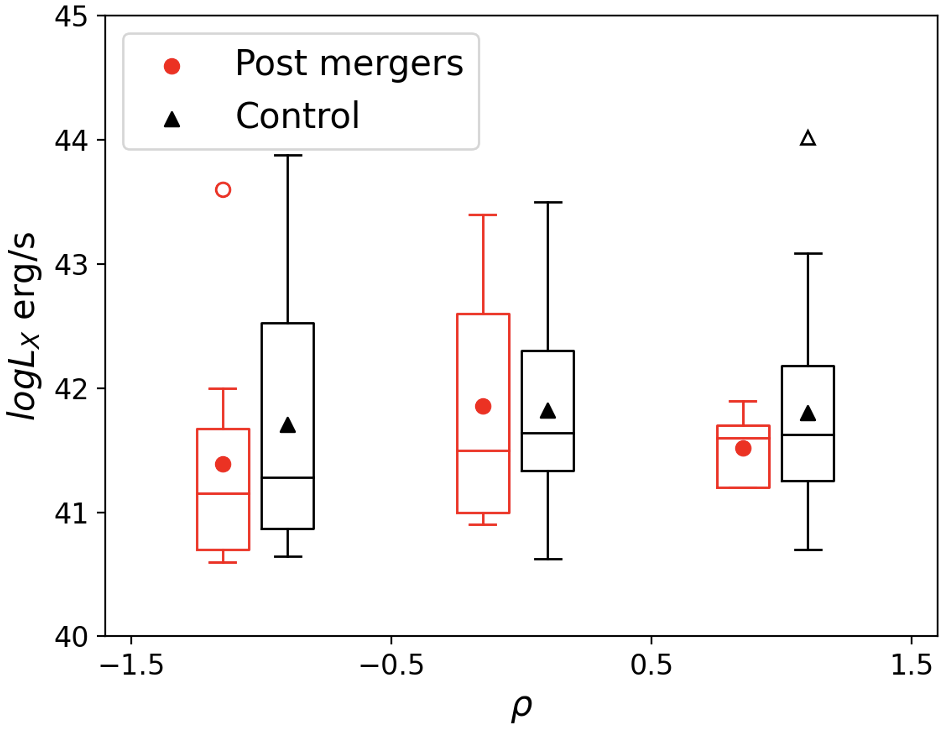}
\end{minipage}
\caption{X-ray luminosity versus (left) stellar mass, (middle) SDSS (g - r) color, and (right) environment density for post-mergers (red filled circles) and control galaxies (black filled triangles). Each plot is divided into three bins. Each box shows the first quartile to the third quartile range of the data in that bin, with a line at the median and a solid symbol at the mean. The whiskers extend from the box by 1.5x the interquartile range. Outliers are shown as open symbols. 
There is no dependence of the mean or median X-ray luminosity of post-mergers or controls on the stellar mass, color, or environment. 
}
\label{fig:Lx_properties}
\end{center} 
\end{figure*}

\subsection{Dependence of the AGN Luminosities on Galaxy Properties}
Figure~\ref{fig:Lx_properties} shows the distribution of the X-ray luminosities of AGN in post-mergers (red circles) and controls (black triangles) in bins of the (left) stellar mass, (middle) color, and (right) environment density. We find no dependence of the AGN luminosity on the stellar mass of the galaxy for both post-mergers and controls. The median X-ray luminosities in both post-mergers and their noninteracting controls are similar. This suggests that while mergers help trigger more AGN, the AGN properties might not depend on the triggering mechanism. There is a large scatter in X-ray luminosities. However, this could be driven by our smaller AGN sample for both post-mergers and control galaxies at large masses.            

The X-ray AGN luminosity of post-mergers also shows no dependence on the galaxy color within errors although there are mild hints that it may increase with blue colors. The control galaxies show no dependence of the AGN luminosity on the galaxy color. Lastly, neither post-mergers nor controls show any dependence of the X-ray luminosity on the environment within our measurement errors. It should be noted that we cannot rule out the presence of strong dust obscuration, which could lead to an underestimate in the measured X-ray luminosities of our sources. This will be investigated in future work with NuSTAR \citep{2013ApJ...770..103H} observations.

\section{Summary}
In this paper, we investigate the relationship between mergers and AGN in the nearby universe with 79 post-merger remnant galaxies and 233 control galaxies with X-ray imaging from Chandra or XMM. 
We find that the X-ray AGN fraction in post-merger galaxies is $\sim 55.7\% \pm 5.6\%$, and it is $\sim$2.4 times higher than that in a mass- and redshift-matched control sample. The significance of this results is $>$6.5$\sigma$. The AGN fraction and excess obtained are sensitive to the AGN selection technique used. An AGN excess is seen using X-rays, optical, and mid-infrared AGN diagnostics but not in the radio. We find that the X-ray AGN fraction of post-merger remnants is independent of the stellar mass and galaxy environment but strongly depends on the galaxy color so that blue galaxies have higher X-ray AGN fractions. On the other hand, the X-ray AGN fraction in noninteracting control galaxies has a strong dependence on the galaxy mass, increasing with the stellar mass, but shows no dependence on the galaxy color or environment given our measurement errors. These trends strongly depend on the AGN wavelength diagnostics used. We find that probing low-luminosity AGN with deep X-ray observations provides a better constraint on the overall AGN fractions. Our work shows that deep X-ray data are essential to understanding the merger–AGN connection. We will be extending this work to lower-mass galaxies with $9<$log M$_*$/M$_{\odot} <10.5$ and investigating quenching signatures in these merger remnants
in future papers.\\

\section*{Acknowledgement}
Support for this work was provided by the National Aeronautics and Space Administration through Chandra Award Number 19700403 issued by the Chandra X-ray Center, which is operated by the Smithsonian Astrophysical Observatory for and on behalf of the National Aeronautics Space Administration under contract NAS8-03060. The scientific results reported in this article are based on observations made by the Chandra X-ray Observatory, data obtained from the Chandra Data Archive and the Chandra Source Catalog 2.0, and software provided by the Chandra X-ray Center (CXC) in the application packages CIAO. 

This work was also based on observations obtained with XMM-Newton, an ESA science mission with instruments and contributions directly funded by ESA Member States and NASA. This research has made use of data obtained from the 4XMM XMM-Newton Serendipitous Source Catalog compiled by the 10 institutes of the XMM-Newton Survey Science Centre selected by ESA.

Coauthor Preethi Nair would also like to acknowledge support by the National Science Foundation under Grant No. 1616547. The NSF grant enabled the visual classification of the parent sample of galaxies used in this analysis. Any opinions, findings, and conclusions or recommendations expressed in this material are those of the author(s) and do not necessarily reflect the views of the National Science Foundation.

This publication makes use of data products from the Wide-field Infrared Survey Explorer, which is a joint project of the University of California, Los Angeles, and the Jet Propulsion Laboratory/California Institute of Technology, funded by the National Aeronautics and Space Administration. This research has made use of ALLWISE catalog from the NASA/IPAC Infrared Science Archive (IRSA), which is funded by the National Aeronautics and Space Administration and operated by the California Institute of Technology.

This publication makes use of the VLA FIRST catalog provided by the National Radio Astronomy Observatory, which is a facility of the National Science Foundation operated under cooperative agreement by Associated Universities, Inc.

This publication makes use of data from the Sloan Digital Sky Survey. Funding for the Sloan Digital Sky Survey has been provided by the Alfred P. Sloan Foundation, the U.S. Department of Energy Office of Science, and the Participating Institutions. SDSS acknowledges support and resources from the Center for High Performance Computing at the University of Utah. The SDSS website is www.sdss.org.

\appendix
\section{Other optical diagnostics}

\begin{figure*}[t!]
\begin{center}
\begin{minipage}{0.95\textwidth}
\includegraphics[width=\linewidth]{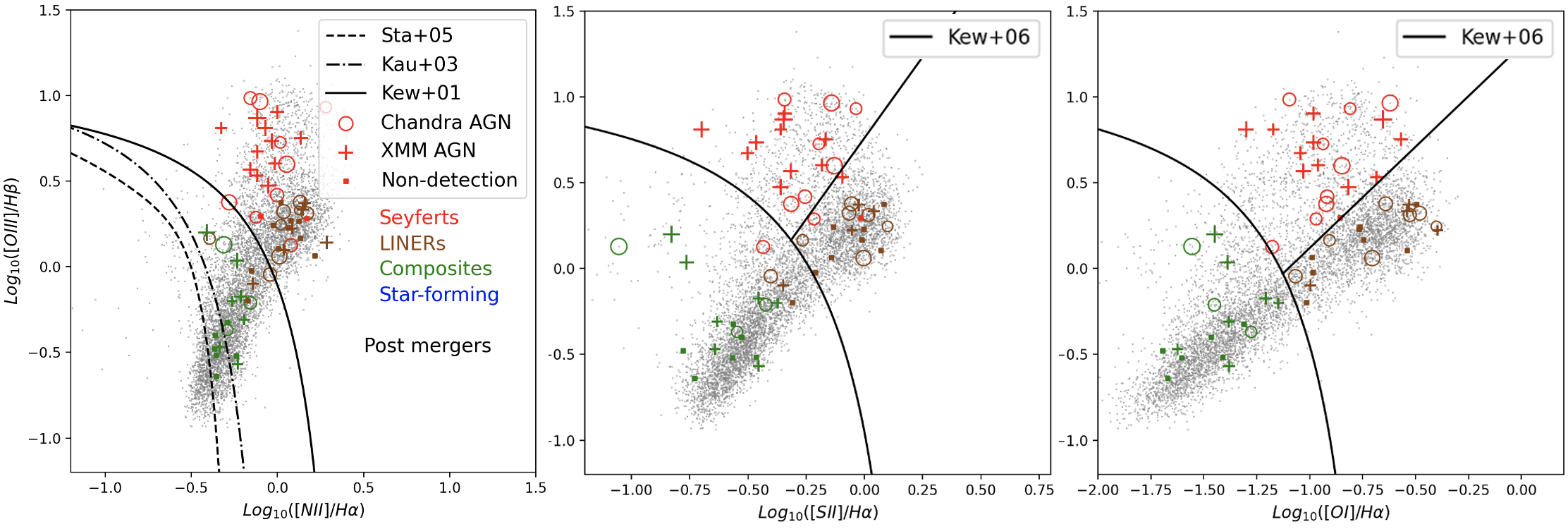}
\end{minipage}
\begin{minipage}{0.95\textwidth}
\includegraphics[width=\linewidth]{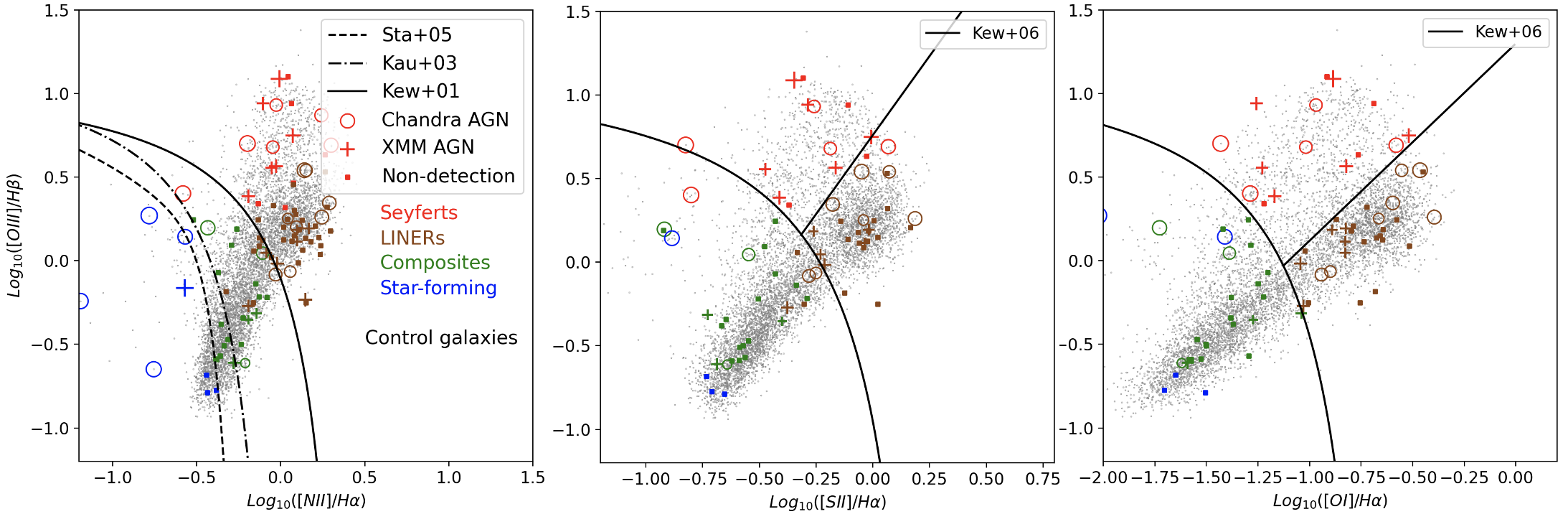}
\end{minipage}
\caption{The [O III]/H$\beta$ vs [N II]/H$\alpha$, [O III]/H$\beta$ vs [S II]/H$\alpha$ and [O III]/H$\beta$ vs [O I]/H$\alpha$ BPT diagrams for post-mergers in the upper panel and control galaxies in the lower panel. The grey points at the background are the general galaxies with log M$_*$/M$_{\odot}>10.5$ in SDSS DR14. Galaxies with an X-ray AGN identified in Chandra are shown with circles. AGN identified in XMM are shown with crosses. Galaxies are classified as Seyferts (red), LINERs (brown), composites (green), and star forming (blue) by the \cite{Stasinska:2006uy} dashed curve and the \cite{2006MNRAS.372..961K} solid lines. The \cite{2003MNRAS.346.1055K} curve is also plotted for comparison.}
\label{fig:optAGNdiagnostics}
\end{center} 
\end{figure*}

The optical AGN diagnostic used in this work is introduced in Section 3, which is the [O III]$\lambda5007$/H$\beta$ vs [N II]$\lambda6584$/H$\alpha$ BPT diagram. We have found that all BPT Seyferts and 45.8\% of LINERs in post-mergers are identified as X-ray AGN by either Chandra or XMM. Here we compare our single BPT diagnostic to other optical AGN diagnostics. 
Figure~\ref{fig:optAGNdiagnostics} shows the combined BPT diagnostics \citep{2006MNRAS.372..961K} for post-mergers in the upper panel and control galaxies in the lower panel. We applied a cut of S/N $> 3$ on all six emission lines ([O III]$\lambda5007$, H$\beta$, [N II]$\lambda6584$, H$\alpha$, [S II]$\lambda6717,6731$, and [O I]$\lambda6300$). 
Galaxies below the \cite{Stasinska:2006uy} dashed curve on the [NII]/H$\alpha$ diagram and below the \cite{2006MNRAS.372..961K} solid curves on [S II]/H$\alpha$ and [O I]/H$\alpha$ diagrams are classified as star forming (blue). 
Galaxies between the dashed curve and the \cite{2001ApJ...556..121K} solid curve on [NII]/H$\alpha$ diagram 
are classified as composites (green). 
Galaxies above either one of the solid curves on these three diagrams are classified as optical AGN (Seyferts + LINERs). 
Seyferts (red) are classified as above either one of the division straight lines on [S II]/H$\alpha$ or [O I]/H$\alpha$ diagrams. 
The 79 post-mergers are classified into 21 Seyferts, 19 LINERs, 16 composites, and no star forming.
The remaining 23 galaxies are unclassified due to a low S/N. Using the combined BPT diagnostics, the optical AGN (Seyferts + LINERs) fraction in post-mergers is $50.6\%\pm5.6\%$. 
The 233 control galaxies are classified into 15 Seyferts, 29 LINERs, 21 composites, and 8 star forming. The remaining 160 galaxies have low S/N and are not classified by the BPT diagnostic. The optical AGN fraction in the control sample is $18.9\%\pm2.6\%$, which implies an optical AGN excess of $\sim 2.7$ in post-mergers relative to control galaxies. 
Applying the equivalent width cut of EWH$\alpha>3$\AA\ , we obtained an optical AGN of $41.8\%\pm5.5\%$ in post-mergers and $12.0\%\pm2.1\%$ in control galaxies. This implies an AGN excess of 3.5$\pm$0.8. 
The AGN excess obtained with this method is higher than that obtained from the single [N II] BPT method. However, this method results in a large unclassified fraction in the sample compared to the single BPT method as we placed S/N cuts on two additional emission lines. 

\begin{figure*}[b!]
\begin{center}
\begin{minipage}{0.47\textwidth}
\includegraphics[width=\linewidth]{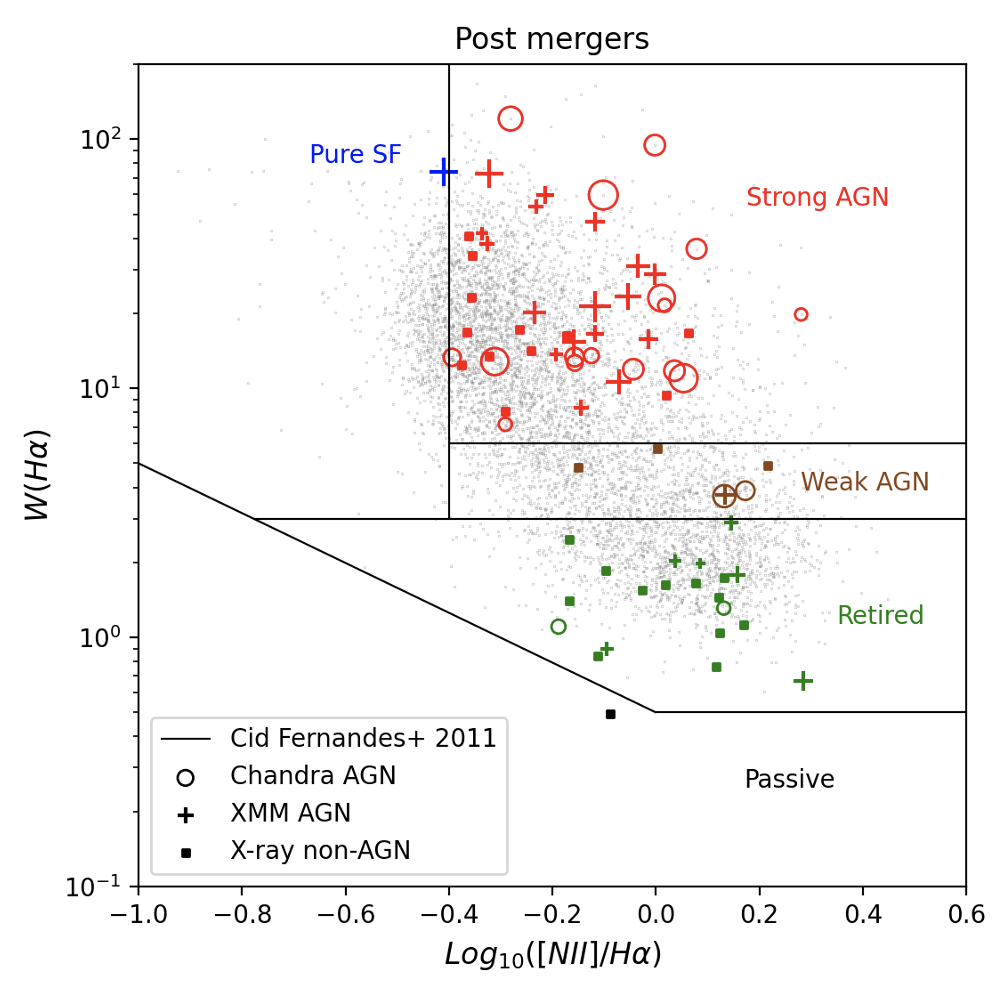}
\end{minipage}
\begin{minipage}{0.47\textwidth}
\includegraphics[width=\linewidth]{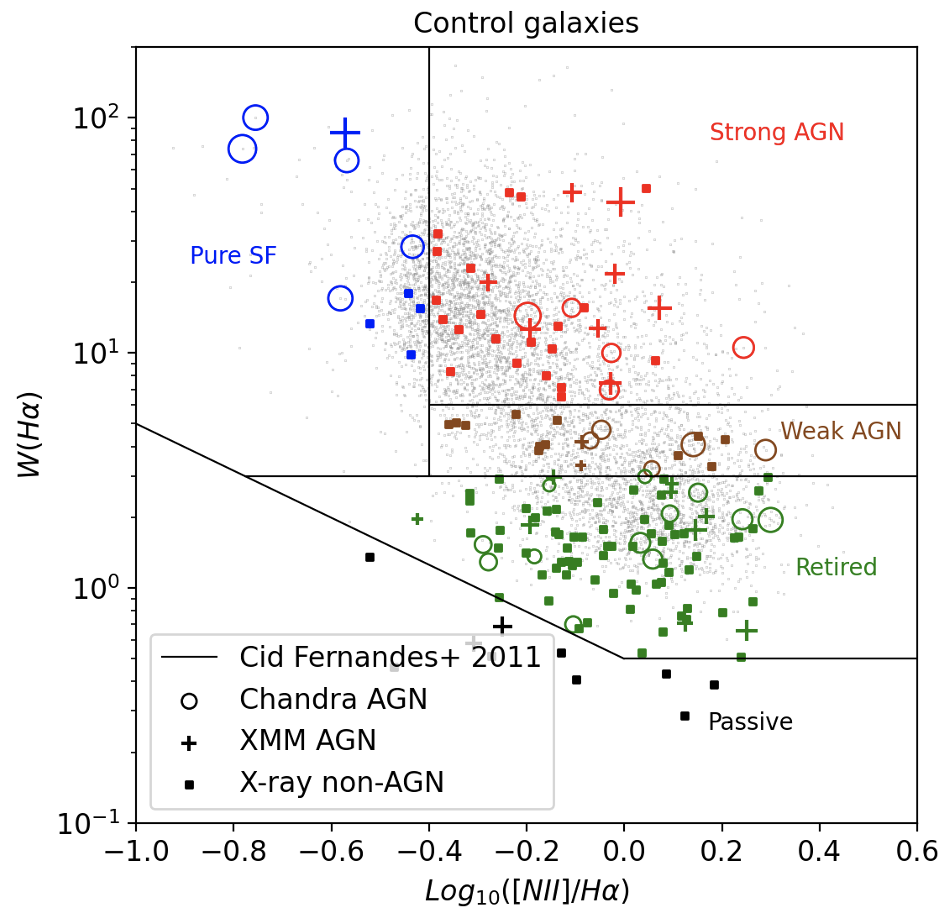}
\end{minipage}
\caption{The H$\alpha$ equivalent width vs [N II]/H$\alpha$ diagrams for post-mergers on the left and control galaxies on the right. The grey points at the background are the general galaxies with log M$_*$/M$_{\odot}>10.5$ in SDSS DR14. Galaxies with an X-ray AGN identified in Chandra are shown with circles. AGN identified in XMM are shown with crosses. Galaxies are classified into Seyferts (red), weak AGN (brown), retired galaxies (green), star-forming galaxies (blue), and passive galaxies (black).}
\label{fig:WHANdiagnostic}
\end{center} 
\end{figure*}

Another optical AGN diagnostic is the WHaN diagram \citep{2011MNRAS.413.1687C}, which uses the H$\alpha$ equivalent width vs [N II]/H$\alpha$ emission line ratio. We applied a cut of S/N $>3$ on H$\alpha$ and [N II]. Pure star-forming galaxies (blue) are classified with log ([N II]/H$\alpha) < -0.4$ and W(H$\alpha$) $>$ 3\AA. Strong AGN (i.e., Seyferts; red) are classified with log ([N II]/H$\alpha) > -0.4$ and W(H$\alpha$) $>$ 6\AA. Weak AGN (brown) are classified with log ([N II]/H$\alpha) > -0.4$ and W(H$\alpha$) between 3 and 6\AA. Retired galaxies (green) are classified with W(H$\alpha$) $<$ 3\AA. Passive galaxies (black) are classified with W(H$\alpha$) and W([N II]) $<$ 0.5\AA. 
LINERs in this method are defined as the combination of weak AGN and retired galaxies. 
Figure~\ref{fig:WHANdiagnostic} shows the WHaN diagnostic for post-mergers (left) and control galaxies (right). 
The 79 post-mergers are classified into 45 strong AGN, 6 weak AGN, 20 retired galaxies, 1 passive galaxy and 1 star-forming galaxy. The remaining 6 galaxies have a low S/N and are unclassified. The optical AGN (strong + weak AGN) fraction in post-mergers with this method is $64.6\%\pm5.4\%$. 
The 233 control galaxies are classified into 34 strong AGN, 19 weak AGN, 87 retired galaxies, 11 star-forming galaxies, and 10 passive galaxies. The remaining 72 galaxies are unclassified due to a low S/N. The optical AGN fraction in the control sample is $22.7\%\pm2.7\%$.
The optical AGN excess is $\sim 2.8$ in post-mergers relative to control galaxies. This is consistent with the optical BPT AGN excess in post-mergers relative to control galaxies. In addition, the number of multiwavelength AGN is consistent no matter which optical diagnostic is used.
However, 38.2\% of the strong AGN classified by the WHaN diagram in post-mergers are X-ray non-AGN, unlike in the [N II]/H$\alpha$ single BPT where all Seyfert post-mergers were X-ray detected. 
This suggests that the WHaN diagram may misclassify some non-AGN host galaxies as strong AGN. 

In summary, we find a strong excess of AGN in post-mergers compared to control galaxies in all optical AGN diagnostics.


\end{document}